\documentclass[journal,10pt]{IEEEtran}

\usepackage{amssymb}
\usepackage{amsmath}
\usepackage{cite}
\usepackage{url}
\usepackage{xcolor}
\usepackage{cite,graphicx,amsmath,amssymb}
\usepackage{epstopdf}
\usepackage{fancyhdr}
\usepackage{mdwmath}
\usepackage{mdwtab}
\usepackage{amsthm}
\usepackage{setspace}
\usepackage{hyperref}
\usepackage{algorithm}
\usepackage{algorithmic}
\usepackage{pdfpages}
\usepackage{mathtools}
\usepackage{makecell}
\hypersetup{colorlinks=false}

\graphicspath{{./src/img/}}

\newtheorem{theorem}{Theorem}

\newtheorem{lemma}{Lemma}

\newtheorem{corollary}{Corollary}

\allowdisplaybreaks
\makeatletter
\def\ScaleIfNeeded{%
\ifdim\Gin@nat@width>\linewidth \linewidth \else \Gin@nat@width
\fi } \makeatother

\begin{document}

\title{Adaptive TTD Configurations for Near-Field Communications: An Unsupervised \\ Transformer Approach}

%\author{
%\IEEEauthorblockN{ Yuanwei~Liu\IEEEauthorrefmark{1}, Zhijin~Qin\IEEEauthorrefmark{1}, Maged Elkashlan\IEEEauthorrefmark{1}, and  Yue~Gao\IEEEauthorrefmark{1}\\} \IEEEauthorblockA{
%\IEEEauthorrefmark{1} Queen Mary University of London, London, UK\\
%%\IEEEauthorrefmark{2} Lancaster University, Lancaster, UK\\
% } }

\author{

% Yuanwei\ Liu, Zhijin\ Qin, Maged\ Elkashlan, Yue\ Gao, and Lajos\ Hanzo

Hsienchih~Ting,
% ~\IEEEmembership{Student Member,~IEEE,}
        % Zhijin~Qin,~\IEEEmembership{Student Member,~IEEE,}
        % Maged~Elkashlan,~\IEEEmembership{Member,~IEEE,}
        Zhaolin~Wang,~\IEEEmembership{Graduate Student Member,~IEEE,}
        and Yuanwei~Liu,~\IEEEmembership{Fellow,~IEEE}

\thanks{The authors are with the School of
Electronic Engineering and Computer Science, Queen Mary University of
London, London E1 4NS, U.K. (email:\{h.ting, zhaolin.wang, yuanwei.liu\}@qmul.ac.uk).}
\vspace{-0.5cm}
}

\maketitle
\begin{abstract}
True-time delayers (TTDs) are popular analog devices for facilitating near-field wideband beamforming subject to the spatial-wideband effect. In this paper, an adaptive TTD configuration is proposed for short-range TTDs. Compared to the existing TTD configurations, the proposed one can effectively combat the spatial-widebandd effect for arbitrary user locations and array shapes with the aid of a switch network. A novel end-to-end deep neural network is proposed to optimize the hybrid beamforming with adaptive TTDs for maximizing spectral efficiency. 1) First, based on the U-Net architecture, a near-field channel learning module (NFC-LM) is proposed for adaptive
beamformer design through extracting the latent channel response features of various users across different frequencies. In the NFC-LM, an improved cross attention (CA) is introduced to further optimize beamformer design by enhancing the latent feature connection between near-field channel and different beamformers. 2) Second, a switch multi-user transformer (S-MT) is proposed to adaptively control the connection between TTDs and phase shifters (PSs). In the S-MT, an improved multi-head attention, namely multi-user attention (MSA), is introduced to optimize the switch network through exploring the latent channel relations among various users. 3) Third, a multi feature cross attention (MCA) is introduced to simultaneously optimize the NFC-LM and S-MT by enhancing the latent feature correlation between beamformers and switch network.
% Two networks, near-field Channel Latent Feature Learning module (NFC-LM) and Switch Multi-user Transformer (S-MT) are proposed to optimize the design of beamformers and adaptivity of TTD connections, respectively. 1) In the NFC-LM, a convolutional encoder is first built to extract the channel latent features of each frequency. Then a cross attention (CA) is proposed to help linear decoder tackle the beamformer design through constructing the relationship between channel latent features and beamformers. 2) In the S-MT, an improved transformer with multi-user attention (MSA) is proposed to adaptively manage the switch through exploring the correlation between spectral efficiency in diverse users and assorted TTD connection situations. Moreover, a multi feature cross attention (MCA) is introduced to simultaneously optimize the NFC-LM and S-MT by establishing the link between beamformers and various connection scenarios. 
Numerical simulation results show that 1) the proposed adaptive TTD configuration effectively eliminates the spatial-wideband effect under uniform linear array (ULA) and uniform circular array (UCA) architectures, and 2) the proposed deep neural network can provide near optimal spectral efficiency, and solve the multi-user bemformer design and dynamical connection problem in real-time.

\end{abstract}

\begin{IEEEkeywords}
{A}daptive true-time delayers, hybrid beamforming, near-field, transformer
\end{IEEEkeywords}

\section{Introduction}
% With the rapid developments of computing chip and artificial intelligence, data-hungry applications such as sending holographic videos and wearing virtual reality devices are likely to become daily applications for people. 
For supporting the enormous data requirements, the sixth generation of wireless technology (6G) in the high-frequency band with ultra-broad bandwidth has drawn extensive attention \cite{THZrefer}. However, the communication distances in high-frequency band, such as millimeter-wave (mmWave) and terahertz (THz) bands, will experience severely decrease \cite{THZdistance}. %Anticipated to play a crucial role in addressing the issues of path loss and distance, extremely large-scale multiple-input multiple-output (XL-MIMO) technology is poised to become a key component in the solution \cite{XLMIMO6G}.
Extremely large-scale multiple-input multiple-output (XL-MIMO) technology, with its potential to effectively address path loss and distance challenges, is thereby emerging as a crucial component in the implementation of 6G \cite{XLMIMO6G}. %While with XL-MIMO and high frequency, 6G not only promises vastly improved data speeds, ultra-low latency, and enhanced reliability, but it also presents several new challenges. 
Although the integration of XL-MIMO and high-frequency technologies promises vastly improved data speeds and ultra-low latency in 6G, they also bring forth several new challenges. \emph{First}, different form the traditional wireless systems where high-frequency communication takes place in the far-field region, the utilization of XL-MIMO can result in high-frequency communication occurring in the near-field region.
% the performance of near-field region bewteen 6G and 5G will show a significant distinction. The utlization of multi bands in the extremely high-frequency results in severe propagation loss within the near-field region. 
As introduced in \cite{XLMIMO6G} and \cite{liu2023near}, the near-field region in 6G can span over hundreds of meters, due to the large aperture of antenna arrays and the extremely high carrier frequencies. Such an extended near-field region in 6G demands a reevaluation of electromagnetic (EM) characteristics, necessitating a shift from conventional modeling methods used in 5G. Particularly, in 6G, the dominance of spherical waves in the near-field region makes the conventional planar-wave model less applicable.
% First, we will face two distinct radiation fields, far-field and near-field. Different from the 5G, the electromagnetic (EM) characteristics of wireless systems need to be reevaluated and utilize multi band in the THz band will come with severe propagation loss in the near-field region. Meanwhile, according to \cite{XLMIMO6G} and \cite{liu2023near} this near-field region can span hundreds of meters due to the extremely large aperture of antenna arrays and the extremely high carrier frequencies. Besides that, in this near-field region we cannot continue using planar waves model to characterize the radiation pattern since the spherical waves become the dominant in the proximity of the antennas. 
\emph{Second}, the increase in both antenna number and bandwidth within high-frequency massive MIMO systems exacerbates the near-field spatial-wideband effect \cite{GaoTTD}. The resulting frequency-dependent variation in array response poses a significant challenge to the conventional phase shifters (PSs) based hybrid beamforming architectures. Current methods \cite{PShyb0,PShyb1,PShyb2} adjust the phase of the signal at each antenna element through generating identical phase shifts across different frequencies, which is inadequate for addressing the near-field spatial-wideband effect in high-frequency massive XL-MIMO systems. 
In high-frequency communication, hybrid beamforming integrates analog and digital processing to enhance signal strength towards the intended receiver with reduced hardware complexity \cite{yan2022dynamic}, overcoming the limitations faced by conventional digital beamforming methods \cite{conventionenergy}.
% In high-frequency band communication, conventional digital beamforming methods are constrained by many hardwares \cite{conventionenergy}. 
% On the contrary, Hybrid beamforming consists of the analog signal processing and the digital processing which can maximize the signal strength in the direction of the intended receiver and also maintain low hardware complexity \cite{yan2022dynamic}. 
% On the contrary, hybrid beamforming, which combines analog signal processing with digital processing, can maximize signal strength towards the intended receiver while maintaining low hardware complexity \cite{yan2022dynamic}. %To address the frequency-dependent spatial-wideband effect in OFDM systems, the use of true-time delayerser (TTD) beamforming methods are be explored, as TTDs can provide a frequency-dependent phase \cite{TTDradar0}, helping to align the phases of the OFDM subcarriers across the bandwidth. 
To address the frequency-dependent spatial-wideband effect in XL-MIMO systems, the implementation of true-time delayers (TTDs) beamforming methods are being explored. TTDs can provide a frequency-dependent phase \cite{TTDradar0}, assisting in the phase alignment of multi subcarriers throughout the bandwidth.
% Based on the conventional Fully-connected (FC) architecture, a direct approach within these hardwares is the substitution of phase shifters (PSs) in conventional hybrid beamforming setups with TTD elements \cite{TTDradar0, TTDradar1, TTDradar2} are proposed. 
Based on the conventional fully-connected (FC) architecture, a straightforward approach involves replacing phase shifters (PSs) with TTDs within the conventional hybrid beamforming framework, as proposed in \cite{TTDradar0, TTDradar1, TTDradar2}. 
% However, this structure will bring unfordable costs of hardware complexity when TTDs tackle the high carrier frequencies. 
However, adopting this structure in an XL-MIMO system will bring unaffordable costs of hardware complexity. In order to reduce hardware complexity, \cite{DPPdai,cui2021near, conventionenergy,GaoTTD,XieDYTTD,najjar2023hybrid,wang2023ttd} proposed to position a limited number of TTDs between the PSs and RF chains to mitigate the spatial-widebandd effect of uniform linear array (ULA). 
% However, a larger array aperture results in a narrower beam, a fixed number of TTDs with predetermined time delays or static connections between TTDs and PSs cannot sustain high beamforming performance. Additionally, the aforementioned method mainly focus on ULA systems and may encounter a significant performance decline when applied to different antenna structures in practical scenarios. Therefore, the main focus of this paper is to explore an effective TTDs based hybrid beamforming structure for arbitrary antenna structures in near-field wideband multi-user systems.
% Therefore, we propose an novel adaptive TTD configuration for near-field beamforming which can configure the time delays and connection between TTDs and PSs adaptively. 

\subsection{Prior Works}
\subsubsection{Conventional Algorithm based Hybrid Beamforming}
TTD provides a notable advantage in maintaining consistent beamforming over a broad frequency range. The frequency-dependent characteristic of TTDs can effectively mitigate the spatial-wideband effect. \cite{GaoTTD} discussed the impact of TTDs in enhancing hybrid beamforming for downlink transmission and integrated TTDs into the PSs-based analog beamformer, offering wideband beamforming capabilities comparable to full-digital arrays. Moreover, the TTD precoding architecture presented in \cite{DPPdai} provides a detailed and structured method, allowing adaption to the system of varying bandwidths by tuning the number of TTDs according to the maximum subcarrier and center frequency ratio. 
% where the TTDs layer is adaptable to systems of varying bandwidths by adjusting the number of TTDs based on the upper bound of the ratio of the maximum subcarrier frequency to the center frequency.
% Moreover, \cite{DPPdai} provided a more detailed and structured approach with its TTD precoding architecture. In this approach the TTDs layer could adapt to systems with different bandwidths since the number of TTDs is designed based on the upper bound of ratio of maximum subcarrier frequency to center frequency.
% In this approach, PSs are divided into several subgroups, ensuring that each RF chain is connected to every antenna through a specific number of TTDs. And these TTDs are only connected to several groups of PSs. 
With similar TTD hybrid beamforming structure, \cite{cui2021near} utilized a piecewise-far-field wideband channel model to approximate the near-field channel model, addressing the near-field beam split challenge in XL-MIMO system. Additionally, the ULA is divided into several small subarrays to manage phase discrepancies in near-field channels by decomposing them into inter-array near-field and intra-array far-field discrepancies. These discrepancies are compensated by PSs and TTDs, respectively.
% They employed a combination of TTDs and PSs to compensate for both far-field and near-field phase discrepancies.
% Moreover, \cite{DPPdai} provided a more detailed and structured approach with its TTD Precoding architecture that PSs are divided into several sub groups, ensuring each RF chain is connected to every antenna and a certain number of TTDs but TTDs are only connected with several PSs groups. 
However, above methods ignore the finite time delay and phase delay constraints of TTDs and PSs, as well as the implications on system size, complexity, and cost. As a balanced solution, a switch network is designed to dynamically control the connection between TTDs and PSs, offering a compromise between system simplicity and spectral efficiency. \cite{conventionenergy} presents the dynamic-subarray with fixed true-time-delay (DS-FTTD) architecture, blending low-cost FTTD elements with a dynamic subarray strategy, but at the cost of reduced spectral efficiency and difficulty in effective beamforming.
% With the FTTD, DS-FTTD can effectively simplify component complexity, but at the cost of reduced spectral efficiency and difficulty in effective beamforming.
% With the similar dynamic switch network structure, 
Similarly, \cite{XieDYTTD} fix the connection between TTDs and PSs while dynamically controlling the connection between the RF chains and TTDs. The proposed hybrid bemforming algorithm optimizes the PSs and TTDs for maximum spectral efficiency, disregarding power consumption, then randomly select the connection between RF chains and TTDs to meet the power constraint. 
% first maximize the spectral effciency by desiging the PSs and TTDs without the power consumption then randomly select the connection between RF chains and TTDs to meat the power constraints. 
% However, implementing this random connection strategy in real-time applications is impractical, as it requires selecting the highest spectral efficiency by traversing all connection cases, based on pre-solved and immutable parameters.
Different from the dynamic connection strategy, \cite{chang2024hybrid} employs baseband TTDs rather than the RF TTDs \cite{DPPdai}, as in this framework they can compute the optimal results of a fully digital structure \cite{el2014spatially} and then apply a linear layer to minimize the mean square error (MSE) between the optimal results and their proposed method.
% Different from the dynamical connection strategy, on the basis of the PS FC structuree, \cite{chang2024hybrid} utlized TTDs to replace the RF chains since under this frame the effect of RF chains can be compensated by TTDs and PSs. And they first calculate the optimal results of fully digital structure \cite{el2014spatially} then employ one linear layer to minimize the mean square error (MSE) between the optimal results and TTDs and PSs. However, the proposed method just consider the far-field planar wave scenario and ignore the severe propagation loss of the near-field spherical model \cite{chen2021hybrid,han2021hybrid}. 
Expanding on the cascaded TTDs structure proposed by \cite{Zhaiserial} for angular coverage expansion, \cite{najjar2023hybrid} tackles the spatial-wideband effect by accumulating time delays. Moreover, \cite{wang2023ttd} explored the advantages of using a serial TTDs configuration over a parallel one in the near-field region. To counteract the reduction in independent control associated with a serial setup, a hybrid configuration is introduced for single-user systems, and for broader coverage in multi-user systems, a hybrid-forward-and-backward (HFB) configuration is proposed.
\subsubsection{Deep Learning based Hybrid Beamforming}
Deep learning (DL) based hybrid beamforming has drawn growing attention. \cite{liyemodeldriven} implemented a model-driven DL approach for hybrid beamforming by incorporating the iterative discrete estimation (IDE2) precoder \cite{model-driven} into a neural network through an unfolding approach. \cite{peken2020deep} proposed a deep neural network (DNN) to learn the SVD process, which is trained by factorizations derived from SVD. Different from above mentioned methods that employ DL to enhance the performance of traditional algorithms, \cite{ElbirDNN} introduced a convolutional neural network (CNN) for direct mapping from input channel matrices to hybrid beamformers, employing supervised learning with a digital codebook as the label.
% nonlinear mapping from the input multi-user channel matrix to the output hybrid beamformers, using supervised learning with a fully digital codebook as the training label. 
Moreover, to address the issue of short-range beamformers design in single user MIMO system, \cite{elbir2019joint} treated antenna selection as a classification task, using two CNNs for joint hybrid beamforming and antenna selection design. 
% Two separate CNNs using channel response as input and employing supervised learning were proposed to jointly design the hybrid beamforming and the connection between PSs and antennas. 
To address the complexities of multi-user MISO system beamformer design, \cite{vu2021machine} integrated a double-loop algorithm with a DNN to accelerate antenna selection, mapping beamforming vectors to configurations more efficiently.  
% Incorporating traditional algorithm, \cite{vu2021machine} a double-loop algorithm for designing beamformers and selecting antennas in multi-user MISO systems. Additionally a DNN is utilized to accelerate the exhaustive search for antenna selection by mapping beamforming vectors to connection scenarios.
However, as the complexity of the system increases, it is hard to transform the problem into a convex problem and calculate the optimal results as training labels. Given the limitations inherent to the MSE framework, the suboptimal training labels produced by traditional algorithms pose a significant challenge for DL models to surpass these conventional methods in performance. In \cite{huang2018unsupervised,lin2019beamforming,hojatian2021unsupervised}, an unsupervised learning approach is employed for MIMO hybrid beamforming, wherein the optimization function and hardware constraints are directly incorporated into the loss function. Furthermore, leveraging a DNN architecture that utilizes the channel matrix as input, the aforementioned DL methods demonstrate the capability to achieve near-optimal outcomes without the necessity for training labels. Moreover, \cite{liu2022deep} uitlized two residual networks \cite{he2016deep} which are trained separately by unsupervised learning to address beamformer design and antenna selection, surpassing traditional algorithms in performance. 
\subsection{Motivation and Contributions}
As mentioned above, TTDs based hybrid beamforming has been widely investigated to combat the spatial-wideband effect in mmWave or THz wireless communication. However, current research mainly focuses on the planar-wave model, an approximation of the spherical-wave, and is often restricted to one specific antenna structure. 
Moreover, given a finite time delay, the independent use of multiple TTDs is hard to effectively alleviate the spatial-wideband effect.
% Besides that, with finite time delay, utilizing multiple TTDs for compensation independently cannot effectively mitigate the spatial-wideband effect. 
Therefore, it is important to introduce an effective TTDs-based hybrid beamforming structure that is adaptable to the spherical-wave model and arbitrary antenna structures.
According to \cite{najjar2023hybrid, wang2023ttd, ZhaoTTD} and \cite{JeongTTD}, serial TTDs configuration can provide adequate time delay for hybrid beamforming with low hardware complexity. However, a larger array aperture results in a narrower beam, a fixed number of TTDs with predetermined time delays or static connections between TTDs and PSs cannot sustain high beamforming performance. Moreover, the complexity of TTDs-based hybrid beamforming system poses computational challenges for conventional algorithms, hindering real-time application capabilities. In addition, supervised learning methods struggle to achieve near-optimal results. Motivated by this, we propose an adaptive TTD configuration hybrid beamforming method for arbitrary antenna structure that cascade multiple TTDs and compensate the time delay adaptively through controlling the connection between TTDs and PSs dynamically. Furthermore, a novel unsupervised network is introduced to optimize the hybrid beamforming with adaptive TTDs for maximizing spectral efficiency.
The main contributions of this paper can be summarized as follows:

\begin{itemize}
  \item  We first introduce an serial TTDs configuration with adaptive switch network for arbitrary user locations and antenna shapes in near-field region. We then propose a U-Net \cite{Unet} structure based near-field channel feature learning module (NFC-LM) for beamformer design. In addition, an improved cross attention (CA) \cite{petit2021u} is proposed to enhance the correlation between the latent features \cite{zhou2021latent} of near-field channel and corresponding beamformers. 
  % CNN structure for beamformer design. A channel feature encoder is utilized to extract the latent feature of wireless channel. In addition, a improved cross attention module is proposed to construct the  relationship among wireless channel, baseband digital beamformer, RF chains , TTDs beamformer and PSs beamformer. 

  \item To realize the adaptive switch network, we further propose a switch multi-user transformer (S-MT) with Hungarian algorithm to adaptively control the connection between the TTDs and PSs. Moreover, in order to effectively enhance the spectral efficiency, we propose an improved multi-head self-attention \cite{vaswani2017attention} called multi-user attention (MSA) to model the channel relationships among various users. 
  % We further propose a dynamic connection structure between TTDs and PSs under serial configuration to enhance the spectral efficiency. For the adaptive TTD configuration, we introduce a multi-user transformer with hungary algorithm to design the connection matrix. The multi-user transformer can be utilized to model the correlations among different users and learn the features of various connection scenarios. 
  
  \item To simultaneously optimize the beamformers design and TTDs connection selection in an end-to-end way, 
  % and to adapt our proposed adaptive TTD configuration to arbitrary antenna shapes, 
  we propose a multi-feature cross-attention (MCA) module further enhancing the latent feature correlations between beamformers and switch network.
  % To jointly optimize the configuration of our adaptive TTD configuration structure, we propose a multi-feature cross-attention block to establish the relationship among connection matrix and different beamformers. Besides that, a multi feature decoder is employed to design the baseband digital beamformer, TTDs beamformer and PSs beamforemer independently.
  \item We provide numerical results for both ULA and UCA architectures to evaluate the performance of our proposed adaptive TTD configuration hybrid beamforming method. The results demonstrate that our method can effectively combat the spatial-widebandd effect for both ULA and UCA, regardless of user locations. In addition, the proposed unsupervised deep neural network can provide near optimal spectral efficiency.
  
\end{itemize}
\subsection{Organization and Notations}
The rest of the paper is organized as follows: Section II intoduces the strucutre of adaptive TTD configuration and formulate the spherical-wave model hybrid beamforming problem. Section III introduces the proposed DL based algorithm for beamformer design and dynamic PSs selection. Section IV presents the numerical results of different array structures and ablation studies of our proposed method. Section V concludes this paper.

\textit{Notations:} We use lower-case, bold-face lower-case and bold-face uppercase letters to represent scalars, vectors and matrices, respectively. The transpose, conjugate transpose of a matrix are denoted by $(\cdot)^T$ and $(\cdot)^H$, respectively. The hadamard product is denoted by \(\odot \). The convolution operate is denoted by \(\ast\). The Euclidean norm of vector \(\mathbf{x}\) is denoted as \( \left \| \mathbf{x} \right \| \), while the Frobenius norm of matrix \(\mathbf{X}\) is denoted as \( \left \| \mathbf{X} \right \|_{F} \). A block diagonal matrix with diagonal blocks \(\mathbf{x}_{1},...,\mathbf{x}_{N}\) is denoted as blkdiag
\(\{\mathbf{x}_{1},...,\mathbf{x}_{N}\}\). \( \mathcal{CN}(\mu, \sigma^2) \) denotes the circularly symmetric complex Gaussian random distribution with mean \(\mu\) and variance \(\sigma^2\). \(\mathcal{I}(M)\) denotes the set \(\{1,2,\ldots, M\}\).
%New expressions are derived in our exact SOP analysis and diversity order analysis.
 %New expressions are derived both by our exact analysis and large scale antenna array analysis.
\begin{figure*}[t!]
\centering
\includegraphics[width=0.6\textwidth]{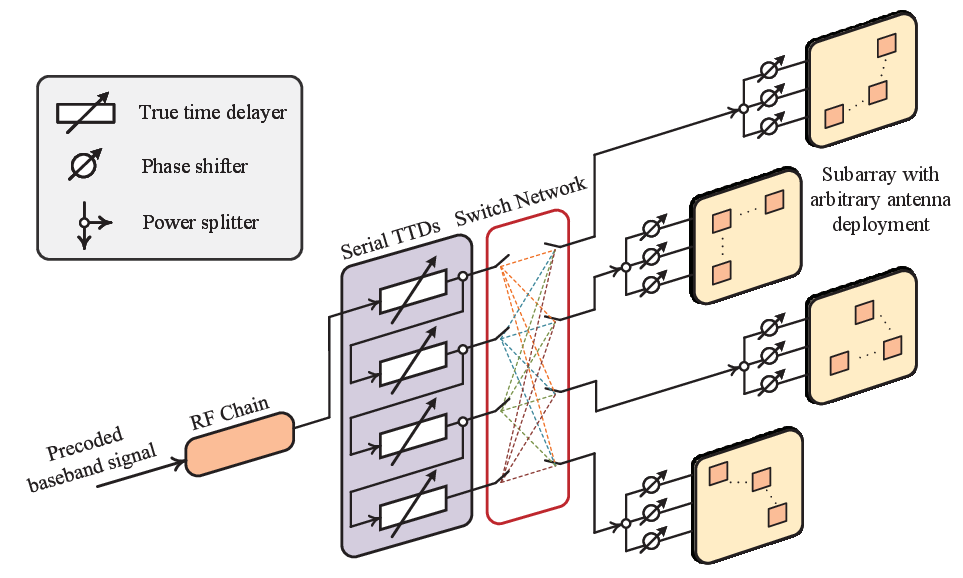}
\vspace{-0.4cm}
\caption{Proposed adaptive-serial configuration for TTD-based hybrid beamforming.}
\label{system_model}
\end{figure*}
\section{System Model and Problem Formulation}
In this paper, we study a near-field wideband XL-MIMO communication system. A base station (BS) equipped with \( N \) antennas serves \( K \) single-antenna users situated in the near-field region. The system operates in the mmWave or THz bands, adopting orthogonal frequency division multiplexing (OFDM) with \(M\)
 subcarriers to address the inter-symbol interference caused by the frequency-wideband effect. The bandwidth and central frequency
 of system are denoted by $B$ and $f_{c}$, respectively,
 therefore each OFDM subcarrier having a frequency \( f_m = f_c + \frac{B(2m-1-M)}{2M} \).
 % A base station is equipped with two distinct types of large-scale antenna array structures: one being a Uniform linear array(ULA) and the other a Uniform circular array(UCA), each containing N antennas. These structures operate in the mmWave or THz bands and are designed to serve K single-antenna users located in the near-field area.
% To counter the frequency-wideband effect caused by
%  adopting a large antenna array and high-frequency band,
%  Orthogonal Frequency Division Multiplexing (OFDM) with M
%  subcarriers is utilized. The bandwidth and central frequency
%  of system are denoted by $B$ and $f_{c}$, respectively,
%  leading to each OFDM subcarrier having a frequency \( f_m = f_c + \frac{B(2m-1-M)}{2M} \).
To facilitate the near-field wideband beamforming with short-range TTDs for arbitrary user locations and array shapes, we propose an adaptive TTDs hybrid beamforming architecture, depicted in Fig. 1. Our proposed architecture introduces an additional switch network which is positioned between the PS network and the TTD network. The adaptive compensation of time delays among different users is achieved through the dynamic interconnection of TTDs and PSs. 
% This process, implemented by the Switch Network, is denoted by \( \mathbf{S} \). In addition, the time delay of each TTD, represented by \( \tau \), is translated into a phase shift of \( e^{-j2\pi f_m \tau} \) for each subcarrier, thereby enabling the frequency-dependent analog beamforming.

 \subsection{Signal Model}
% We assume that \( N_{\mathrm{R}} \) RF chains feed \( L \) TTDs. Subsequently, these \( L \) TTDs are connected to \( N \) PSs, which in turn are connected to \( N \) antennas. The \( N \) Phase Shifters (PSs) are allocated into \( P \) clusters, where \(P = L\). Each cluster comprises \( Q = N/P \) PSs and is connected to a corresponding TTD.
 % Let $t_{n,l}$ denote the time delay introduced by the $l$-th TTD connected to the $n$-th RF chain.
Let $N_{\mathrm{RF}}$ denote the number of RF chains employed in the adaptive TTDs beamforming architecture. For the $m$-th subcarrier, let $\mathbf{A}_m \in \mathbb{C}^{N \times N_{\mathrm{RF}}}$ denotes the frequency-dependent analog beamformer jointly achieved by PSs and TTDs, and $\mathbf{D}_m \in \mathbb{C}^{N_{\mathrm{RF}} \times K}$ denote the baseband digital beamformer for $K$ users. Then, the transmit signal can be expressed as 
\vspace{-0.5cm}
\begin{equation}
    \mathbf{x}_m = \mathbf{A}_m \mathbf{D}_m \mathbf{c}_m = \sum_{k=1}^K \mathbf{A}_m \mathbf{d}_{m,k} c_{m,k},
    \vspace{-0.3cm}
\end{equation}

where $\mathbf{d}_{m,k} \in \mathbb{C}^{N_{\mathrm{RF}} \times 1}$ is the vector at the $k$-th column of matrix $\mathbf{D}_m$ and also denotes the baseband digital beamformer for the $k$-th user. Vector $\mathbf{c}_m = [c_{m,1},...,c_{m,K}]^T \in \mathbb{C}^{K \times 1}$ denote the unit-power information symbols delivered to the $K$ users, which are modelly as independent and identically distributed (i.i.d.) random variables, i.e., $\mathbb{E}[\mathbf{c}_m \mathbf{c}_m^H] = \mathbf{I}_{K}$. Accordingly, the signal received at user $k$ on the $m$-th subcarrier can be modelled as
\vspace{-0.1cm}
\begin{equation} 
\label{receive_siganl}
    y_{m,k} = \mathbf{h}_{m,k}^H \mathbf{A}_m \mathbf{d}_{m,k} c_{m,k} + \sum_{i \neq k} \mathbf{h}_{m,k}^H \mathbf{A}_m \mathbf{d}_{m,i} c_{m,i} + z_{m,k},
\end{equation}
where $\mathbf{h}_{m,k} \in \mathbb{C}^{N \times 1}$ denotes the baseband communication channel for user $k$ on the $m$-th subcarrier and $z_{m,k} \sim \mathcal{CN}(0, \sigma_{m,k}^2)$ denotes the additive complex Gaussian noise with power of $\sigma_{m,k}^2$.  

In this paper, the spherical-wave-based near-field channel model is considered. The adopted model accounts for the effects of both direct line-of-sight (LOS) and $L_k$ indirect non-line-of-sight (NLOS) paths for user $k$, which arise due to scattering. Hence, the channel \(\mathbf{h}_{m,k}\) can be expressed as \cite{near-fieldYW}
\vspace{-0.2cm}
\begin{equation}
    \mathbf{h}_{m,k} = \beta_{m,k}\mathbf{b}^{*}(f_{m},r_{k},\theta_{k}) + \sum_{l=1}^{L_{k}} \tilde{\beta}_{m,k,l}\mathbf{b}^{*}(f_{m},\tilde{r}_{k,l},\tilde{\theta}_{k,l}),
    \label{eq:hm,k}
\end{equation}
where $\beta_{m,k}$ and $\tilde{\beta}_{m,k}$ denote the complex channel gain for LOS and NLOS paths, respectively, $r_k$ and $\theta_k$ denote the distance and direction of user $k$ with respect to the BS, $\tilde{r}_{k,l}$ and $\tilde{\theta}_{k,l}$ denote the distance and direction of the $l$-th resolvable scatter in user $k$'s NLOS path with respect to the BS, and vector $\mathbf{b}^{*}(f, r, \theta) \in \mathbb{C}^{N \times 1}$ denotes the array response at distance $r$ and direction $\theta$. Due to the spherical wave propagation in the near-field region, $\mathbf{b}^{*}(f, r, \theta)$ should be modelled accurately as 
\vspace{-0.2cm}
\begin{equation}
    \mathbf{b}(f,r,\theta) = e^{-j \frac{2\pi f}{c}\mathbf{r}(r,\theta)},
    \label{brtheta}
    \vspace{-0.2cm}
\end{equation}
where $\mathbf{r}(r,\theta) = [r_1, ..., r_N]^T$
denotes the vector of distances from each antenna to a given point, where $r_n$ specifies the distance from the $n$-th antenna and the location point $(r, \theta)$.
% with $r_n$ denoting the distance between the $n$-th antenna and the location point $(r, \theta)$. 
There are two key observations from the expression of array response. On the one hand, the array response vector is frequency-dependent, which thus requires exploiting TTDs to facilitate frequency-dependent analog beamforming. On the other hand, the distance vector $\mathbf{r}(r,\theta)$ not only depends on $r$ and $\theta$, but also depends on the location of the antenna elements. In other words, different shapes of antenna arrays can lead to different channel characteristics. Based on these observations, in the following, we propose an adaptive TTD configuration to compensate for the spatial wideband effect regarding different array shapes and short-range TTDs. 

\subsection{Proposed Adaptive-Serial TTD Configuration}
% In this subsection, we present the specific structure of the analog beamforming matrix $\mathbf{A}_m$ in the proposed adaptive-serial TTD configuration. 
In this subsection, we introduce the detailed configuration of the analog beamforming matrix, $\mathbf{A}_m$, within the structure of the proposed adaptive-serial TTDs beamforming method. As shown in Fig. \ref{system_model}, we assume each RF chain connects to all antennas via \(L\) TTDs and \(N\) PSs. Furthermore, each TTD is connected to a sub-array with size $Q = N/L$. Let $t_{l,i}$ and $\phi_{n,i}$ denote the time delay at the output of the $l$-th TTD and the phase adjustment of the $n$-th PS connected to the $i$-th RF chain.  Given that PSs only allow constant-modulus, the PSs are subject to the following constraint
\vspace{-0.2cm}
\begin{equation}
\left| \phi_{n,i} \right| = 1.
\vspace{-0.2cm}
\label{eq:modulephi}
\end{equation}
To address the short-range limitation, the TTDs are connected in an serial manner \cite{wang2023ttd}. More particularly, the output time delay $t_{l,i}$ of $l$-th TTD is an accumulative result containing all time delays provided by previous TTDs. Hence, $t_{l,i}$ can be expressed as 
\vspace{-0.2cm}
\begin{equation}
    t_{l,i} = \sum_{j=1}^l \tilde{t}_{j,i},
    \vspace{-0.2cm}
\end{equation}
where $\tilde{t}_{j,i}$ denotes the time delay realized by the $j$-th TTD and is subject to a maximum delay constraint, i.e., $\tilde{t}_{j,i} \in [0, t_{\max}]$.
For each subcarrier indexed by \( m \), the frequency domain phase shift realized by the cumulative time delay \( t_{l,i} \) is given by \( e^{-j2\pi f_{m} t_{l,i}} \). We further assume that power is distributed equitably across all TTDs, with adjustments to the power being made through the coefficients of the power divider. Additionally, the dynamic connection between the $l$-th TTD and the $n$-th PS is implemented by the switch network through the adjustment of the binary switch coefficient $\tilde{s}_{n,l,i} \in \{0,1\}$. Consequently, with this architecture, the overall analog adaptive TTD-based hybrid beamforming matrix \( \mathbf{A}_{m} \in \mathbb{C}^{N \times N_{\mathrm{RF}}} \) can be expressed as
% Consequently, with this architecture, the matrix \( \mathbf{A}_{m} \in \mathbb{C}^{N \times N_{RF}} \) which characterizes the aggregate analog beamformer on the \(m\)-th subcarrier for all \(N_{RF}\) RF pathways can be expressed accordingly as \\
% \(\mathbf{F}_{m} = \frac{1}{\sqrt{N}} \times \)
\vspace{-0.2cm}
\begin{equation}
\mathbf{A}_{m} = \begin{bmatrix}
a_{m,1,1} & \cdots & a_{m,1,N_{\mathrm{RF}}} \\
\vdots & \ddots & \vdots \\
a_{m,N,1} & \cdots & a_{m,N, N_{\mathrm{RF}}}
\end{bmatrix},
\vspace{-0.1cm}
\end{equation}
% \begin{equation}
% \scalebox{0.7}{ % Adjust the scale factor as needed to fit the matrix within the page
% $ % Start math mode
% \begin{bmatrix}
% \mathbf{\varphi}_{1,1} c_{1,1} e^{-j2\pi f_{m} \tau_{1,1}} & \cdots & \mathbf{\varphi}_{N_{RF},1} c_{N_{RF},1} e^{-j2\pi f_{m} \tau_{N_{RF},1}} \\
% \vdots & \ddots & \vdots \\
% \mathbf{\varphi}_{1,N_{TTD}} c_{1,N_{TTD}} e^{-j2\pi f_{m} \tau_{1,N_{TTD}}} & \cdots & \mathbf{\varphi}_{N_{RF},N_{TTD}} c_{N_{RF},N_{TTD}} e^{-j2\pi f_{m} \tau_{N_{RF},N_{TTD}}}
% \end{bmatrix}
% $ % End math mode
% }
% \end{equation}
where 
% \begin{align*}
%     \mathbf{F}_m[(l-1)Q+n,r] = f_{m,r,l,n} = \varphi_{r,l,n} s_{r,l,n} e^{-j 2\pi f_m \tau_{r, l}}.
% \end{align*}
\begin{equation}
    a_{m,n,i} = \phi_{n,i} \sum_{l=1}^L \tilde{s}_{n,l,i} e^{-j 2\pi f_m t_{l,i}}.
    \vspace{-0.2cm}
\end{equation}
To make the adaptive-serial TTD configuration viable in practical systems, it is important to ensure that each PS is connected exclusively to one corresponding TTD. This constraint can be expressed as 
\vspace{-0.2cm}
\begin{equation}
    \sum_{l=1}^{L} \tilde{s}_{n,l,i} = 1. \label{eq:wc}
    \vspace{-0.2cm}
\end{equation}
Furthermore, since each TTD in a serial configuration can support a larger time delay compared to a parallel configuration \cite{wang2023ttd}, we distribute the $N$ antennas into $L$ groups. Each TTD then exclusively connects to one sub-array with size $Q = N/L$. In this case, we have:
\vspace{-0.2cm}
\begin{equation}
    \sum_{n=1}^{N} \tilde{s}_{n,l,i} = Q.\label{eq:wr} 
    \vspace{-0.2cm}
\end{equation}

We note that based on the above configuration, if we consider $N$ antennas with different PSs should partitioned into $L$ different TTDs and each TTD at least has one connection, then the total number of combinations can be expressed by the Stirling number of the second kind \cite{graham1989concrete}
\vspace{-0.2cm}
\begin{equation}
    L!\times S(N,L) = L!\times\frac{1}{L!} \sum_{i=0}^{L} (-1)^{L-i} \binom{L}{i} (i)^{N}.
    \label{eq:SN}
    \vspace{-0.2cm}
\end{equation}
This number is substantial even for a modest number of TTDs. For instance, with a single RF chain, the count reaches \( 3.4032 \times 10^{38} \) for a configuration of 64 antennas and 4 TTDs. Moreover, if we assume that each TTD is connected to non-ordered, equal-sized subsets of antennas, with each subset comprising \( Q = N/L \) antennas, as written in (\ref{eq:wr}) then the total number of combinations, calculated as \( \frac{N!}{\left(N/L!\right)^L L!} \), remains significantly large. For instance, in a configuration with with 64 antennas and 4 TTDs, this leads to approximately \( 2.67 \times 10^{34} \) combinations. To find the near optimal solution, we methodically preallocate the antennas into \(L\) organized groups, with each group containing \(Q\) antennas, as shown in Fig. \ref{system_model}. In this case, the overall analog beamforming matrix $\mathbf{A}_m$ reduces to 
\begin{equation}
\mathbf{A}_{m} = \begin{bmatrix}
\mathbf{a}_{m,1,1} & \cdots & \mathbf{a}_{m,1,N_{\mathrm{RF}}} \\
\vdots & \ddots & \vdots \\
\mathbf{a}_{m,L,1} & \cdots & \mathbf{a}_{m,L, N_{\mathrm{RF}}}
\end{bmatrix}.
\vspace{-0.2cm}
\end{equation}
Here, $\mathbf{a}_{m,p,i} \in \mathbb{C}^{Q \times 1}$ denotes the analog beamformer for the $p$-th ordered, equal-sized subarray and the $i$-th RF chain. It can be written specifically as 
\vspace{-0.2cm}
\begin{equation}
    \mathbf{a}_{m,p,i} = \boldsymbol{\phi}_{p,i} \sum_{l=1}^L s_{p,l,i} e^{-j 2\pi f_m t_{l,i}}, 
    \vspace{-0.2cm}
\end{equation}
where $\boldsymbol{\phi}_{p,i} = [ \phi_{(p-1)Q+1, i},...,\phi_{pQ, i} ]^T$ denotes the PS-based analog beamformer for the $p$-th ordered, equal-sized subarray and the $i$-th RF chain, and $s_{p,l,i} \in \{0,1\}$ is the new binary switch coefficient. In particular, $s_{p,l,i} = 1$ implies that the $l$-th TTD is connected to the $p$-th sub-array with the $i$-th RF chain. Furthermore, we assume that each TTD is exclusively linked to a single sub-array, and vice vesa, each sub-array is uniquely connected to one TTD, resulting in the following constraint:
\vspace{-0.2cm}
\begin{equation}
    \sum_{l=1}^L s_{p,l,i} = 1, \quad \sum_{p=1}^L s_{p,l,i} = 1.
    \vspace{-0.4cm}
\end{equation}

\subsection{Problem Formulation}
In this work, we aim to maximize spectral efficiency by jointly optimizing the analog and digital beamformers. According to \eqref{receive_siganl}, the achievable rate for user $k$ at the $m$-th subcarrier can be calculated based on the Shannon formula as follows:
\begin{equation}
    R_{m,k} = \log_2 \left(1 + \frac{|\mathbf{h}_{m,k}^H \mathbf{A}_m \mathbf{d}_{m,k}|^2}{\sum_{i=1, i \neq k}^K |\mathbf{h}_{m,k}^H \mathbf{A}_m \mathbf{d}_{m,i}|^2 + \sigma_{m,k}^2} \right).
\end{equation}
The spectral efficiency of the considered multi-user OFDM system is thus given by 
\vspace{-0.2cm}
\begin{equation}
    R = \frac{1}{M + L_{\mathrm{CP}}} \sum_{m=1}^M \sum_{k=1}^K R_{m,k}.
    \vspace{-0.2cm}
\end{equation}
The spectral efficiency maximization problem is thus given by 
\vspace{-0.4cm}
\begin{subequations}
    \begin{align}
        &\max_{\mathbf{\Phi}, \mathbf{S}, \mathbf{T}, \mathbf{D}_m}  \quad \sum_{m=1}^M \sum_{k=1}^K R_{m,k} \label{eq:maxRm1}\\
        \vspace{-0.1cm}
        \mathrm{s.t.} \quad & \|\mathbf{A}_m \mathbf{D}_m \|_F^2 \le P_t, \forall m, \\
        \vspace{-0.1cm}
        & \mathbf{a}_{m,p,i} = \boldsymbol{\phi}_{p,i} \sum_{l=1}^L s_{p,l,i} e^{-j 2\pi f_m t_{l,i}}, \forall m, p, i, \\
        \vspace{-0.1cm}
        & \sum_{l=1}^L s_{p,l,i} = 1, \quad \sum_{p=1}^L s_{p,l, i} = 1, \forall i, \\
        & t_{l,i} = \sum_{j=1}^l \tilde{t}_{j,i}, \quad \tilde{t}_{j,i} \in [0, t_{\max}], \forall l,i, \\
        & \left| \phi_{n,i} \right| = 1, \forall n,i,
        \vspace{-0.4cm}
    \end{align}
\end{subequations}
where $P_t$ denotes the maximum transmit power for each sub-carrier. Matrices $\mathbf{\Phi} \in \mathbb{C}^{L \times N_{\mathrm{RF}}}$, $\mathbf{S} \in \mathbb{C}^{L \times L \times N_{\mathrm{RF}}}$, and $\mathbf{T} \in \mathbb{C}^{L \times N_{\mathrm{RF}}}$ represent the coefficients of PSs, switches, and TTDs, respectively. Their entries are given by 
\vspace{-0.2cm}
\begin{equation}
    [\mathbf{\Phi}]_{p,i} = \phi_{p,i}, \quad [\mathbf{S}]_{p,l,i} = s_{p,l,i}, \quad  [\mathbf{T}]_{l,i} = t_{l,i}.
    \vspace{-0.2cm}
\end{equation}

However, jointly optimizing \(\mathbf{\Phi}\), \(\mathbf{S}\), \(\mathbf{T}\) and \(\mathbf{D}_{m}\) is intractable, as \(\mathbf{S}\) directly influences the structure of \(R_{m,k}\) in (\ref{eq:maxRm1}). The aim of this paper is to develop a solution for problem (\ref{eq:maxRm1}) that is not only low in complexity but also feasible for real-time implementation. To enhance the clarity of this paper, we introduce an end-to-end deep learning (DL) approach and a more detailed discussion will be provided in Section \ref{se:DLFRAME}.

\section{DL-Based Adaptive TTD Hybrid Beamforming}\label{se:DLFRAME}

To address the aforementioned problems, in this section, we will introduce our proposed DL-Based Adaptive TTD Hybrid Beamforming algorithm including the network structure, training procedure, and the loss function design. We break down optimization problem into two phases. In the first phase, we design a near-field channel learning module (NFC-LM) to construct the hybrid beamforming matrices \(\mathbf{\Phi}\), \(\mathbf{T}\), and \(\mathbf{D}_m\). In the second phase, we introduce the switch multi-user transformer (S-MT) module to manage the connection matrix \(\mathbf{S}\). These two networks are then trained in an unsupervised end-to-end way.

% Subsequent sections focus on problem (13) in the context of fixed Time Delay Transmission (TTD) connection frameworks for multiple users. This discussion aims to clarify the basic principles of designing wireless channel cross-attention based Encoder-Decoder Network for configurations of PS beamformers, TTD beamformers and baseband digital beamformers. Following this, the text explores the development of multi-user Transformer-based adaptive TTD connection scenarios.

\subsection{Near-Field Channel Latent Feature Learning}

\begin{figure} [t!]
\centering
\includegraphics[width= 3.4in, height=2.30in]{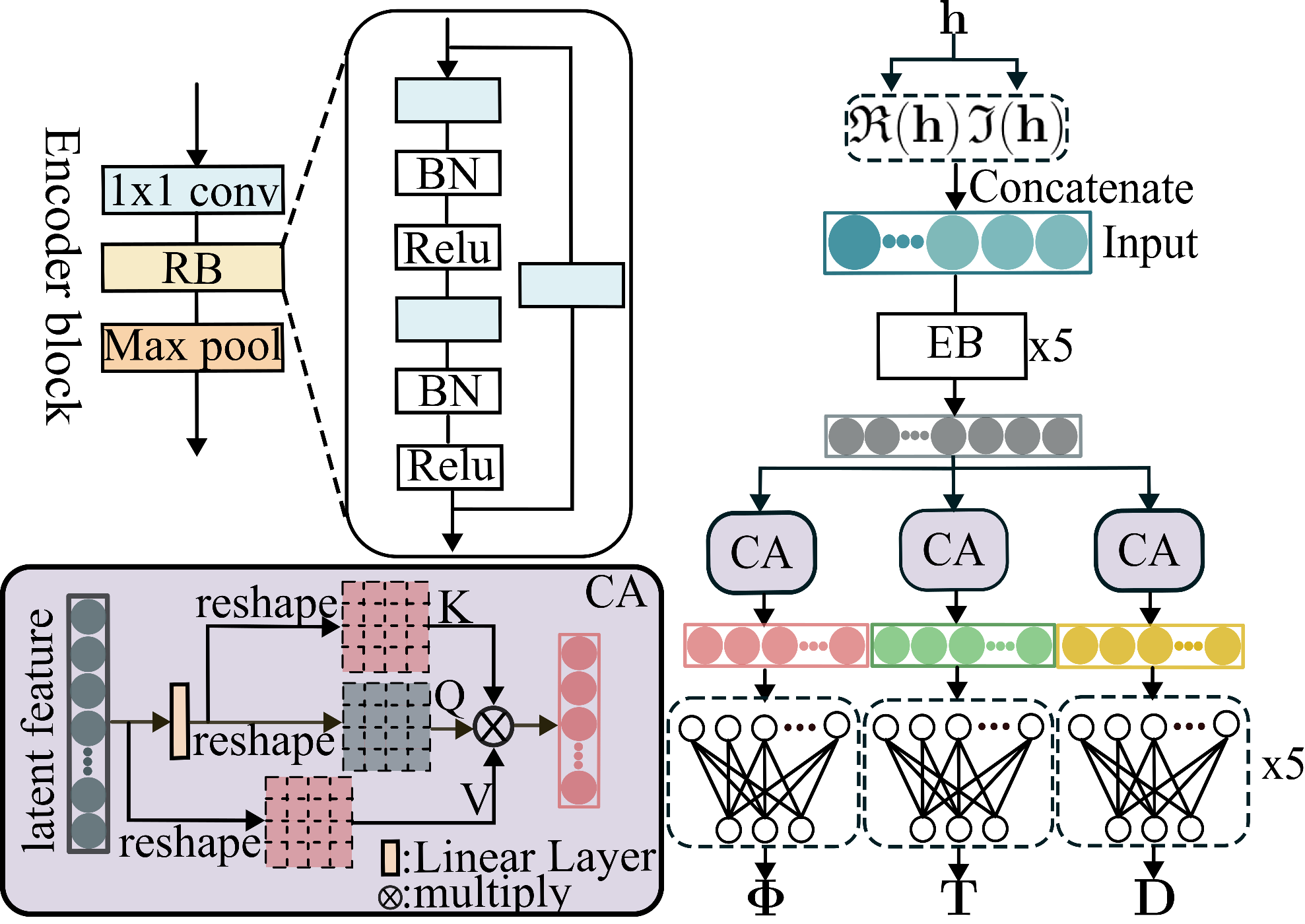}
\vspace{-0.3cm}
 \caption{The network structure of channel feature learning module.
  }
 \label{img:NFC-LM}
\end{figure}

Unlike most existing NN-based hybrid beamforming algorithms \cite{huang2018unsupervised,lin2019beamforming,hojatian2021unsupervised,chang2024hybrid}, which typically employ several linear layers to directly construct relationships among channel response, analog beamformers and digital beamformers for specific antenna shape, our approach introduces a novel unsupervised Encoder-Decoder structure for arbitrary antenna shapes. This structure aims to explore latent feature relations among \(\mathbf{h}_{m,k}\), \(\mathbf{\Phi}\), \(\mathbf{T}\) and \(\mathbf{D}_m\). Based on the U-Net structure \cite{Unet}, our near-field channel learning module (NFC-LM) is divided into two symmetrical parts: a convolutional encoder and multi-linear decoders, as illustrated in Fig. \ref{img:NFC-LM}. Each convolution of convolutional encoder can be expressed as
\vspace{-0.2cm}
\begin{equation}
\mathrm{Conv}_{i,j}(\mathbf{X}) =\mathbf{W}_{i,j} \ast \mathbf{X} + \mathbf{B}_{i,j},
\label{conv}
\vspace{-0.2cm}
\end{equation}
where \(\mathbf{X}\) represents the input tensor, \(\mathbf{W}_{i,j}\) and \(\mathbf{B}_{i,j}\) represents the weights and biases of the \(j\)-th 1D convolution of \(i\)-th convolutional layer, respectively. 
Each linear layer of multi-linear decoders can be expressed as 
\vspace{-0.2cm}
\begin{equation}
\mathrm{Linear}_{i,j}(\mathbf{X}) = \mathbf{W}_{i,j} \mathbf{X} + \mathbf{B}_{i,j},
\label{linear}
\vspace{-0.2cm}
\end{equation}
where \(\mathbf{X}\) represents the input tensor, \(\mathbf{W}_{i,j}\) and \(\mathbf{B}_{i,j}\) represents the weights and biases of \(j\)-th linear projection of \(i\)-th linear decoder layer, respectively.
To enhance latent feature representation learning in NFC-LM, we initially separate complex channel response matrices into their real and imaginary components. These components are treated as multi-modal features, which are then transformed into tensors. Thus, the channel response \(\mathbf{h}_{m,k}\) can be further expressed as follow:
% To enhance latent feature representation learning in NFC-LM, we decompose complex channel response matrices into their real and imaginary parts, treating these as multi-modal features, and then tensorize them. Thus, the channel response \(\mathbf{h}_{m,k}\) can be further expressed as follow:
% To facilitate the latent feature learning of NFL-LM, we decompose the complex channel response matrices into the real domain and image domain as multi modal feature, then vectorized. To do this, we first express the (\ref{eq:hm,k}) as
\vspace{-0.2cm}
\begin{align}
\left[ \begin{array}{c}
\mathfrak{R}(\mathbf{h}_{m,k}) \\
\mathfrak{I}(\mathbf{h}_{m,k})
\end{array} \right]
&= 
\left[ \begin{array}{c}
\mathfrak{R}(\beta_{m,k}) \\
\mathfrak{I}(\beta_{m,k})
\end{array} \right] \odot 
\left[ \begin{array}{c}
\mathfrak{R}(\mathbf{b}^{*}_{m,k}) \\
\mathfrak{I}(\mathbf{b}^{*}_{m,k})
\end{array} \right] \\
+& \sum_{l=1}^{L_{k}}
\left[ \begin{array}{c}
\mathfrak{R}(\tilde{\beta}_{m,k,l}) \\
\mathfrak{I}(\tilde{\beta}_{m,k,l})
\end{array} \right] \odot 
\left[ \begin{array}{c}
\mathfrak{R}(\mathbf{b}^{*}_{m,k,l}) \\
\mathfrak{I}(\mathbf{b}^{*}_{m,k,l})
\end{array} \right],
\label{eq:RIhm,k}
\vspace{-0.2cm}
\end{align}
where \(\mathfrak{R}(\cdot)\) and \(\mathfrak{I}(\cdot)\) denote the real and imaginary parts, respectively. Subsequently, the real and imaginary components of the channel responses across different users and frequencies are combined and transformed into a tensor representation. The process can be expressed as follows:
\begin{equation}
\mathbf{H} = \mathrm{Tens}(\mathrm{con}(\mathfrak{R}(\mathbf{h}_{1,1}), \mathfrak{I}(\mathbf{h}_{1,1}),\ldots,\mathfrak{R}(\mathbf{h}_{M,K}), \mathfrak{I}(\mathbf{h}_{M,K}))),
\label{eq:VecConcat}
\end{equation}
where \(\mathbf{H} \in \mathbb{R}^{K \times M \times 2N}\) represents the combined responses for all users \(K\) and frequencies \(M\), expanding each antenna response into \(2N\) elements, thereby accommodating both its real and imaginary components. Here \(\mathrm{Tens}(\cdot)\) represents the tensorization operation flattening a matrix into a tensor by column order and \(\mathrm{con}(\cdot)\) denotes the concatenation operation.
Additionally, to directly construct the latent feature correlations among channel response, PSs, TTDs and baseband digital beamformer, we also predict the real and imaginary parts of \(\mathbf{\Phi} \in \mathbb{C}^{L\times{N_{RF}}}\), \(\mathbf{T}\in \mathbb{C}^{L \times N_{RF}}\) and \(\mathbf{D}\in \mathbb{C}^{K\times M \times N_{RF}}\). 
Here \(\mathbf{D}\) denotes the concatenation of \(\mathbf{D}_m\) across all subcarriers and all RF chains. 
% Here \(\mathbf{T}\) and \(\mathbf{D}\) denotes the concatenation of \(\mathbf{T}_m\) and \(\mathbf{D}_m\) across all subcarriers and all RF chains, respectively. 
By denoting the input channel response tensor as \(\mathbf{H}\) and the output as \(\mathbf{\Phi}\), \(\mathbf{T}\) and \(\mathbf{B}\), the end-to-end relationship of the NFC-LM can be expressed as
\vspace{-0.1cm}
\begin{equation}
\big[\mathfrak{R}(\mathbf{\Phi}), \mathfrak{I}(\mathbf{\Phi}), \mathfrak{R}(\mathbf{T}), \mathfrak{I}(\mathbf{T}), \mathfrak{R}(\mathbf{D} ), \mathfrak{I}(\mathbf{D} )\big] = \mathrm{NFC\textendash LM}(\mathbf{H}),
\end{equation}
where \(\mathrm{NFC\textendash LM}(\cdot)\) represents the mapping function of NFC-LM.

As showed in Fig. \ref{img:NFC-LM}, the feature extraction part of our NFC-LM consists of five encoder blocks denoted as \(\mathrm{EB}(\cdot)\), which are integrated with batch normalization (\(\mathrm{BN}(\cdot)\)) and \(\mathrm{ReLU}(\cdot)\) activation. Each encoder block includes a 1D convolution layer with a kernel size of 3 and a stride of 1, followed by a residual block denoted as \(\mathrm{RB}(\cdot)\), before the max pooling operation. Considering the input channel response tensor, \(\mathbf{H}\), the feature extraction process for the output latent channel response feature , \(\mathbf{L}_{\mathbf{H}}\), from the encoder can be written as follows:
\vspace{-0.1cm}
\begin{equation}
\begin{split}
\mathrm{RB}_{i} (\mathbf{X}_{i}) &= \mathrm{Conv}_{i,3}(\mathrm{ReLU}((\mathrm{Conv}_{i,2} (\mathbf{X}_{i})))) + \mathbf{X}_{i}, \\
\mathrm{EB}_{i} (\mathbf{X}_{i}) &= \mathrm{Conv}_{i,4}(\mathrm{RB}_{i}(\mathrm{ReLU}(\mathrm{BN}(\mathrm{Conv}_{i,1}(\mathbf{X}_{i})))),\\
\mathbf{L}_{\mathbf{H},i} &= \mathrm{EB}_{i}(\mathbf{L}_{\mathbf{H},i-1}),
\end{split}
\label{eq:bblatent}
\vspace{-0.2cm}
\end{equation}
where \(i\) indexes the \(i\)-th encoder block with being 5 in this paper, \(\mathbf{X}_{i}\) represents the input tensor to the \(i\)-th encoder block, the initial input \(\mathbf{L}_{\mathbf{H},0}\) is set to \(\mathbf{\mathbf{H}}\) and \(\mathbf{L}_{\mathbf{H}_,i}\in \mathbb{R}^{(K\cdot2^{2+i}) \times (M/2^{i}) \times (2N/2^{i})}\). We also introduce an improved cross-attention (CA) \cite{petit2021u} module in NFC-LM to further enhance the latent feature representation. The extracted latent channel response feature \(\mathbf{L}_{\mathbf{H},5}\) is first flattened into \(\mathbb{R}^{K\cdot2^{2+5} \times (M/2^{5})(2N/2^{5})} \),  where \(K\cdot2^{2+5}\) represents the channel dimension and \((M/2^{5})(2N/2^{5})\) reprensents the data length. In the CA module, the process begins with employing a linear layer to map \(\mathbf{L}_{\mathbf{H},5}\) onto the latent features associated with the various beamformers, namely \(\mathbf{L}_{\mathbf{\Phi}}\), \(\mathbf{L}_{\mathbf{T}}\) and \(\mathbf{L}_{\mathbf{D}}\). After that, three distinct linear layers are applied to generate the query (\(\mathbf{Q}\)) from latent channel response feature and key (\(\mathbf{K}\)) and value (\(\mathbf{V}\)) from corresponding latent beamformer feature. Subsequently, the cross-attention weights are calculated by applying the softmax function (\(\mathrm{softmax(\cdot)}\)) to the scaled dot-product of \(\mathbf{Q}\) and \(\mathbf{K}\). This calculation aims to model the interrelations between the latent channel response feature and corresponding latent beamformer feature across the data dimension. Finally, the cross-attention weights are multiplied with \(\mathbf{V}\) to further enhance the connection between channel response and corresponding beamformer. The incorporation of the CA module enable beamformers to tackle arbitrary user locations and antenna shapes through establishing data dimension connections between latent channel response feature and latent beamformer features. To clarify, the operation of the Cross-Attention (CA) module, denoted as (\(\mathrm{CA}(\cdot)\)) can be expressed as follows:
\vspace{-0.2cm}
\begin{subequations}
   \begin{align}
    \mathbf{L}_{\mathbf{Y}} &= \mathrm{Linear}_{0,1}(\mathbf{L}_{\mathbf{X}}), \\
    \mathbf{Q} &= \mathrm{Linear}_{0,2}(\mathbf{L}_{\mathbf{X}}),\\ \label{eq:CAQ}
    \mathbf{K} &= \mathrm{Linear}_{0,3}(\mathbf{L}_{\mathbf{Y}}),\\ \label{eq:CAK}
    \mathbf{V} &= \mathrm{Linear}_{0,4}(\mathbf{L}_{\mathbf{Y}}),\\ \label{eq:CAV}
    \mathbf{A} &= \mathrm{softmax} (\frac{\mathbf{Q}\mathbf{K}^{T}}{\sqrt{d_{k}}}), \\ \label{eq:CAA}
    \mathbf{L}_{\mathbf{Y}} &= \mathbf{A} \cdot \mathbf{V},
\end{align} 
\label{eq:CAAV}
\end{subequations} where \(\mathbf{L}_{\mathbf{X}} = \mathbf{L}_{\mathbf{H}_5}\) represents the input latent feature of CA module, \(\mathbf{L}_{\mathbf{Y}} \in \{\mathbf{L}_{\mathbf{\Phi}}, \mathbf{L}_{\mathbf{T}}, \mathbf{L}_{\mathbf{D}} \}\) represents the resulting latent beamformer feature from CA module, and \(d_k\) represents the dimensionality of key, used for scaling the dot products. Considering the input \(\mathbf{L}_{\mathbf{H},5}\) and the output \(\mathbf{L}_{\mathbf{\Phi}}\), \(\mathbf{L}_{\mathbf{T}}\) and \(\mathbf{L}_{\mathbf{D}}\) of CA module (\(\mathrm{CA}(\cdot)\)) can be expressed as follows
\vspace{-0.2cm}
\begin{subequations}
   \begin{align}
    \mathbf{L}_{\mathbf{\Phi}} = \mathrm{CA}(\mathbf{L}_{\mathbf{H},5}), \mathbf{L}_{\mathbf{T}} = \mathrm{CA}(\mathbf{L}_{\mathbf{H},5}), \mathbf{L}_{\mathbf{D}} = \mathrm{CA}(\mathbf{L}_{\mathbf{H},5}),
    \vspace{-0.2cm}
\end{align} 
\end{subequations}
where each invocation of the \(\mathrm{CA}(\cdot)\) applies different linear projections tailored to the specific beamformer features it generates.
Consequently, the outputs of the CA module are then fed into the multi-linear decoders, denoted as \(\mathrm{MLD}(\cdot)\). These linear decoders mirror the convolutional encoder structure, each comprising five linear layers, and the decoding process can be described as follow
\begin{equation}
\begin{split}
    \mathrm{MLD}(\mathbf{L}_{\mathbf{X}}) &= \mathrm{Linear}_{1,1}(\mathrm{Linear}_{2,1}(\cdots\mathrm{Linear}_{5,1}(\mathbf{L}_{\mathbf{X}}))),\\
    \mathbf{Y} &= \mathrm{MLD}(\mathbf{L}_{\mathbf{X}}),
\end{split}
\end{equation}
where \(\mathbf{L}_{\mathbf{X}} \in \{ \mathbf{L}_{\mathbf{\Phi}}, \mathbf{L}_{\mathbf{T}}, \mathbf{L}_{\mathbf{D}} \}\) is the input of multi-linear decoders and \(\mathbf{Y} = [\mathfrak{R}(\mathbf{\Phi}), \mathfrak{I}(\mathbf{\Phi}), \mathfrak{R}(\mathbf{T}), \mathfrak{I}(\mathbf{T}), \mathfrak{R}(\mathbf{D} ), \mathfrak{I}(\mathbf{D} )]\) is the output of multi-linear decoders.

\vspace{-0.2cm}
\subsection{Switch Multi-User Transformer}
\begin{figure} [t!]
\centering
\includegraphics[width= 3.4in, height=3.4in]{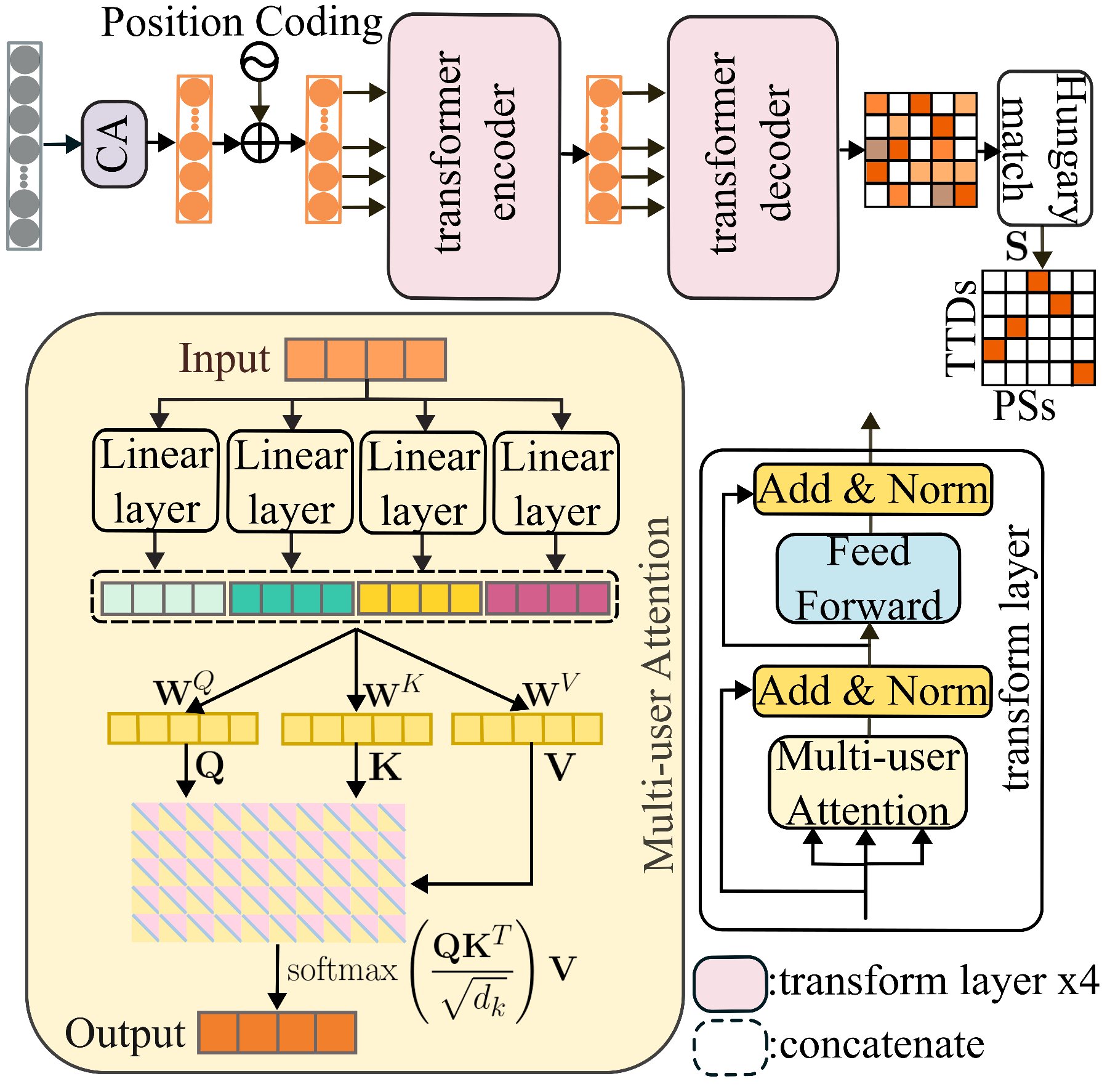}
\vspace{-0.4cm}
 \caption{The network structure of multi-user transformer for adaptive connection between TTDs and PSs.
  }
 \label{MulTrans}
\end{figure}

Our research focuses on near-field hybrid beamforming, particularly addressing the challenge posed by the diverse range of antenna shapes encountered in practical applications. This diversity highlights the necessity for an adaptive TTD configuration approach to ensure effective beamforming across various scenarios.
% Our primary focus is on near-field hybrid beamforming, where the variability of antenna shapes in practical applications necessitates the exploration of an adaptive TTD configuration method. 
 Motivated by this, we introduce the switch multi-user transformer (S-MT), which is designed for beamforming of arbitrary antenna shapes. Specifically, as illustrated in Fig. \ref{system_model}, the S-MT enable the switch network to dynamically control the compensation for time delays through effectively managing the connection between TTDs and PSs. Here, we preallocate the antennas into equal-sized subarrays and construct the correlations between TTD and subarray instead of individual antennas. Thereby, we can reduce the complexity of extremely large antenna modeling and tackle beamforming for arbitrary antenna shapes. As illustrated in Fig.~\ref{MulTrans}, the S-MT comprises a positional encoding, an multi-head attention mechanism, an encoder-decoder transformer, and the Hungarian matching algorithm, which collectively determine the final connection configuration.
 % The switch network, as illustrated in Fig. \ref{system_model}, positioned between the PSs and TTDs, dynamically controls the compensation for time delay, is realized by the S-MT. 

To establish the latent feature interrelation between switch network and channel response, we first employ the CA module to generate the latent switch feature, denoted as \(\mathbf{L_{S}} \in \mathbb{R}^{N_{RF}L \times N_{RF}L}\). In the real-time beamforming process, accurately capturing the intrarealtions among various subarrays is essential. To this end, we assign a unique positional encoding, denoted as $\mathbf{P_C}$, to each PS group to differentiate and model these subarrays effectively. Following the generation of the latent switch feature, represented as \(\mathbf{L_{S}}\),  we proceed by applying a specific positional encoding function, denoted as \(\mathrm{E}_{pc}(\cdot)\). Each PS cluster
% and to facilitate the adaptation of the S-MT to arbitrary antenna shapes, 
is uniquely characterized through a sinusoidal coding approach \cite{vaswani2017attention}, which can be expressed as follows:
\vspace{-0.2cm}
\begin{subequations}
\begin{align}
&\mathrm{E}_{PC}(p,2i) = \sin(p / 1000^{2i/d}), \\
&\mathrm{E}_{PC}(p,2i+1) = \cos(p / 1000^{2i/d}), \\
&\mathbf{P_C} = \mathrm{E}_{PC}(p), \forall p \in \mathcal{I}(N_{RF}L)
\end{align}
\label{eq:pc}
\end{subequations}
where \(\mathbf{P_C} \in \mathbb{R}^{N_{RF}L\times N_{RF}L}\) represents the unique positional encoding for each PS cluster generated by \(\mathbf{E}_{pc}(\cdot)\), and \(p = {1, 2, \dots, N_{RF}L}\) represents the index of preallocated subarray, for example \(p = 1\) indicating the first PS cluster, and so on. Here, \(i\) denotes the coding dimension, and \(d = L = 128\) represents the input feature dimension. Given that the dimensions of \(\mathbf{P_C}\) align with those of the input latent switch feature $\mathbf{L_S}$, we can directly integrate the \(\mathbf{P_C}\) with \(\mathbf{L_{S}} \), which can be written as follow
\vspace{-0.2cm}
\begin{equation}
\mathbf{L_{S}} = \mathbf{L_{S}} + \mathbf{P_C},
\label{eq:pc_update}
\vspace{-0.2cm}
\end{equation}
where \(\mathbf{P_C}\) is calculated as previously described \eqref{eq:pc}.

The encoded latent switch feature \(\mathbf{L_{S}}\) is then fed into a transformer encoder, which comprises \(I\) distinct transformer layers. Each transformer layer is structured with a multi-user self-attention (\(\mathrm{MSA}(\cdot)\)) block, a residual learning block, a layer normalization (\(\mathrm{LN}(\cdot)\)), and a feed-forward network (\(\mathrm{FFN}(\cdot)\)), as depicted in Fig.~\ref{MulTrans}.
In order to facilitate the modeling of different user beamforming characteristics across various subspaces, multiple linear layers are employed in the  \(\mathrm{MSA}(\cdot)\) to generate distinct latent switch feature for different users. 
% The \(\mathrm{MSA}(\cdot)\) block first utilize multiple linear layers to generate the latent feature maps of different users, then employs multiple self-attention layers operating in parallel on the input features to help the model capture characteristics of different users across various subspaces. 
The input to the \(\mathrm{MSA}(\cdot)\) is formulated as follows:
\vspace{-0.2cm}
\begin{equation}
\begin{split}
{\mathbf{L_{S}}}' = \mathrm{con}\Big(\mathrm{Linear}_{0,5}({\mathbf{L_{S}}} ),&\mathrm{Linear}_{0,6}({\mathbf{L_{S}}} ), \cdots\\
&,\mathrm{Linear}_{0,5+K}({\mathbf{L_{S}}} ) \Big),
\label{eq:QKVSELF0}
\end{split}
\vspace{-0.2cm}
\end{equation}
where \(\mathbf{L_{S}}' \in \mathbb{R}^{KN_{RF}L\times N_{RF}L}\) represents the concatenated latent switch feature maps for different users. The output of a single self-attention (\(\mathrm{SA}(\cdot)\)) layer within the \(\mathrm{MSA}(\cdot)\) block is expressed as
\vspace{-0.2cm}
\begin{subequations}
\label{eq:QKVSELF1}
\begin{gather}
\mathbf{Q} = \mathrm{Linear}_{0,K+6} (\mathbf{L_{S}}'), \mathbf{K} = \mathrm{Linear}_{0,K+7} (\mathbf{L_{S}}'), \\
\mathbf{V} = \mathrm{Linear}_{0,K+8} (\mathbf{L_{S}}'), \mathbf{A} = \mathrm{softmax} \left(\frac{\mathbf{Q}\mathbf{K}^{T}}{\sqrt{d_{k}}}\right),\\
\tilde{\mathbf{L_{S}}} = \mathbf{A} \cdot \mathbf{V},
\end{gather}
\end{subequations}
\noindent where \(\mathbf{A}\) represents the calculated self-attention weights, signifying the varying beamforming effects among users. The \(\mathrm{MSA}(\cdot)\) further enhance the  relative importance among users by aggregating outputs from several self-attention layers, described as
\vspace{-0.2cm}
\begin{equation}
\begin{split}
\mathrm{MSA}(\mathbf{L_{S}}') = \mathrm{Linear}_{0,K+9}\Big(\mathrm{con}\big(\mathrm{SA}&(\mathbf{L_{S}}')_1, \mathrm{SA}(\mathbf{L_{S}}')_2,  \\
&\ldots, \mathrm{SA}(\mathbf{L_{S}}')_J\big)\Big).
\label{eq:MSA}
\end{split}
\end{equation}
Here, \(J=K\) represents the number of users, and \(\mathrm{SA}(\mathbf{L_{S}}')_j\) denotes the output from the \(j\)-th parallel self-attention layer. The latent switch feature maps resulting from the \(\mathrm{MSA}(\cdot)\) layer are subsequently passed to the \(\mathrm{FFN}(\cdot)\).
% Furthermore, the encoded sequence \(L^{\prime}_{w}\) is
% forwarded to the transformer encoder structure which has \(L\) transformer layers. Each layer consists of multi-user self attention (MSA) block, a residual learning block, a layer normalization (LN), and a feed forward network (FFN), as showed in Fig. \ref{MulTrans}. In the MSA, multiple attenion layers are applied to the input feature in parallel, so that the model can capture feature of different users among different subspaces. Similarly as the equation (\ref{eq:QKV}) and (\ref{eq:ATN}), the output can be calculated as
% \begin{equation}
% \begin{aligned}
% Q &= W^Q L_w, \\
% K &= W^K L_w, \\
% V &= W^V L_w, \\
% L_w^{\prime} &= ATN \cdot V,
% \end{aligned}
% \label{eq:QKVSELF}
% \tag{25}
% \end{equation}
% where \(ATN\) is the calculated attention weight represent the varying importance of current users. With the single attention layer, the MSA is formed by concatenating the result of several attention layers, which can be represented by
% \begin{equation}
% \begin{aligned}
% \text{Head}_i &= L_{w_i}^{\prime}, \\
% L_w^{\prime} &= \text{con}[\text{Head}_1, \text{Head}_2, \ldots, \text{Head}_h]W,
% \end{aligned}
% \label{eq:MSA}
% \tag{26}
% \end{equation}
% where i is the number of users, the term \(\text{Head}_i\) denotes the \(i^{th}\) result of the parallel attention layer.The output feature map of the MSA layer is then passed to the feedforward layer.
To clarify the computational process within the \(i\)-{th} transformer layer ((\(i \in [1,2,...,I=8]\))), the encoding-decoding process is expressed as follows:
\vspace{-0.2cm}
\begin{align}
 \mathrm{FFN}(\mathbf{X}) &= \mathrm{Linear}_{i,2}(\mathrm{GELU}(\mathrm{Linear}_{i,3}(\mathbf{X}))), \\
\tilde{\mathbf{L_{S}}}_i &= \mathrm{MSA}(\mathrm{LN}(\tilde{\mathbf{L_{S}}}_{i-1})) + \tilde{\mathbf{L_{S}}}_{i-1}, \\
\tilde{\mathbf{L_{S}}}_{i} &= \mathrm{FFN}(\tilde{\mathbf{L_{S}}}_{i}) + \tilde{\mathbf{L_{S}}}_{i},
\end{align}
where \(\mathbf{X}\) represents the input feature to the \(\mathrm{FFN(\cdot)}\), \(\tilde{\mathbf{L_{S}}}_{i}\) represents the feature processed by the \(i\)-th transformer layer, with \(\tilde{\mathbf{L_{S}}}_{0} = \mathbf{L_{S}}\) being the initial input latent switch feature, and \(\mathbf{S}^{'} = \tilde{\mathbf{L_{S}}}_{8} \in \mathbb{R}^{L\times L}\) indicating the predicted connection matrix of the final output. The \(\mathrm{FFN}(\cdot)\) block consists of two linear layers with a Gaussian Error Linear Unit (\(\mathrm{GELU}\)) activation function in between. The input and output feature dimensions for the \(\mathrm{FFN}(\cdot)\) are set to 128, and the dimension of the intermediate layer is expanded to 512 to enhance feature representation ability.
% The dimension of input and output features in FNN is 512 while the feature dimension of linear projection layer is 1024 to enhance the representation. Specifically, we propose a improved multi head attention mechanism which called multi user attention to tackle the adaptive beamformer connection among different users. First, we employ K(represents the number of users) linear projection layers to the input \(L_{C}\) to model the variable connection situations of different users. 

% \begin{equation}
% L_{C_{k}} = LN_{k}(L_C), k\in[1,K]
% \end{equation}

% Then, we concatenate the multi user connection feature representations\(L_{C} = \mathrm{Concatenate}(L_{C_1},L_{C_2},....,L_{C_K})\). After that, we still utilize the self-attention based method to construct the correlation among different users which can be expressed as:
% \begin{equation}
% L_{C} = L_{C} \cdot \mathrm{Softmax}(\frac{L_{C}\times L^{T}_{C}}{\sqrt{D}})
% \end{equation}
% where D is the dimension size of \(L_{C}\) and in this paper D is 512. 
% symmetrically, we construct the transformer decoder.
Considering the extremely large number of subarrays, it becomes impractical to directly converts the continuous values in the predicted connection map \(\mathbf{S}^{'} \in \mathbb{R}^{N_{RF}L\times N_{RF}L}\) into a binary matrix through classification methods \cite{elbir2019joint}. Additionally, each subarray must be uniquely connected to one TTD, and vice versa. To effectively tackle this challenge, the Hungatian algorithm is employed to optimize the connection selection of \(\mathbf{S}^{'}\). We initially define a permutation matrix \(\mathbf{P} \in \mathbb{R}^{N_{RF}L\times N_{RF}L}\), which indicates the assignments of subarrays to TTDs. The algorithm treats \(\mathbf{S}^{'}\) as the initial cost matrix, denoted as \(\mathbf{C} \in \mathbb{R}^{N_{RF}L\times N_{RF}L}\).
% it is necessary to select an appropriate \(\mathbf{S} \in \mathbb{Z}^{P\times P}\) from \(\mathbf{S}^{'}\). However, translating \(\mathbf{S}^{'}\) with its continuous predicted values into a binary matrix is challenging. Directly treating this as a classification problem becomes impractical with the presence of a large number of PSs and TTDs. In practical scenarios, each PS must be assigned to exactly one TTD, and each TTD must connect to exactly one PS. To address this, we apply the Hungarian algorithm to guide the selection of \(\mathbf{S}\). Initially, we define \(\mathbf{P} \in \mathbb{R}^{P \times P}\) as the permutation matrix representing the assignment of PSs to TTDs. We consider \(\mathbf{S}^{\prime}\) as the initial cost matrix \(\mathbf{C} \in \mathbb{R}^{N_{RF}L \times N_{RF}L}\). 
Then the problem is reformulated to maximize the sum of products between the assignment decisions and their corresponding costs. The algorithm starts by subtracting the minimum value from each row and column of their respective elements, resulting in a matrix with zero-valued elements indicative of potential assignments. Through iterative refinement, we arrive at an optimal pairing between TTDs and PSs, as illustrated in \textbf{Algorithm 1}.

% given a predicted connection assessment map \(\tilde{C}\), the assignment problem can be formulated as:
% \begin{subequations}
%     \begin{align}
%     & \max_{P} \sum_{i=1}^{n} \sum_{j=1}^{n} P_{ij} \cdot \tilde{C}_{ij}\\
%     \textbf{Subject to:} \\
%         & \sum_{j=1}^{n} P_{ij} = 1,\forall i \in \{1, \ldots, n\} \quad \label{eq:agent} \\
%         & \sum_{i=1}^{n} P_{ij} = 1,\forall j \in \{1, \ldots, n\} \quad  \label{eq:task} \\
%         & P_{ij} \in \{0, 1\},\forall i,j \in \{1, \ldots, n\} \quad  \label{eq:binary}
%     \end{align}
% \end{subequations}

% Here, \(P\) is is the permutation matrix representing the assignment of PSs to TTDs and \(n\) is the number of TTDs and PSs(in this paper both equals \(N_{TTD}\)). The objective is to maximize the sum of the products of the assignment decisions and the corresponding costs, ensuring that each PS is assigned to exactly one TTD and each TTD is given exactly one PS. The decision variable \(P_{ij}\) is binary, indicating whether task \(j\) is assigned to agent \(i\)(1) or not (0). However, \(\tilde{C}\) with continuous predicted value is hard to project into a binary matrix. And it is unrealistic tohandle this problem as a classification problem directly when there are massive antennas and TTDs. Therefore, we apply the Hungarian algorithm to project the output connection assessment map \(\tilde{C}\) to a binary connection matrix.

\begin{algorithm}[t!]
\caption{Hungarian Algorithm for Assignment Problems}
\begin{algorithmic}[1]
\STATE Let $\mathbf{C}$ be the initial cost matrix with dimensions $N_{RF}L \times N_{RF}L$
% \State $\tilde{C} \leftarrow C$ after performing row and column reductions
\STATE The objective is to maximize the total cost of the assignment:
\STATE $\underset{\mathbf{P}}{\max} \sum_{i=1}^{n} \sum_{j=1}^{n} P_{ij} \cdot C_{ij}$
\FOR{$i \in \{1, \ldots, n\}$}
    \STATE $C_{i,:} \leftarrow C_{i,:} - \min(C_{i,:})$
\ENDFOR
\FOR{$j \in \{1, \ldots, n\}$}
    \STATE $C_{:,j} \leftarrow C_{:,j} - \min(C_{:,j})$
\ENDFOR
\WHILE{number of lines $P < n$}
    \STATE $s \leftarrow \min(\text{uncovered elements of } \mathbf{C})$
    \STATE Adjust \(\mathbf{C}\) with respect to $s$
\ENDWHILE
\STATE Derive $\mathbf{P}$ from \(\mathbf{C}\) as the assignment matrix
\RETURN $\mathbf{P}$
\end{algorithmic}
\end{algorithm}

\begin{figure*}[t!]
\centering
\includegraphics[width=0.7\textwidth]{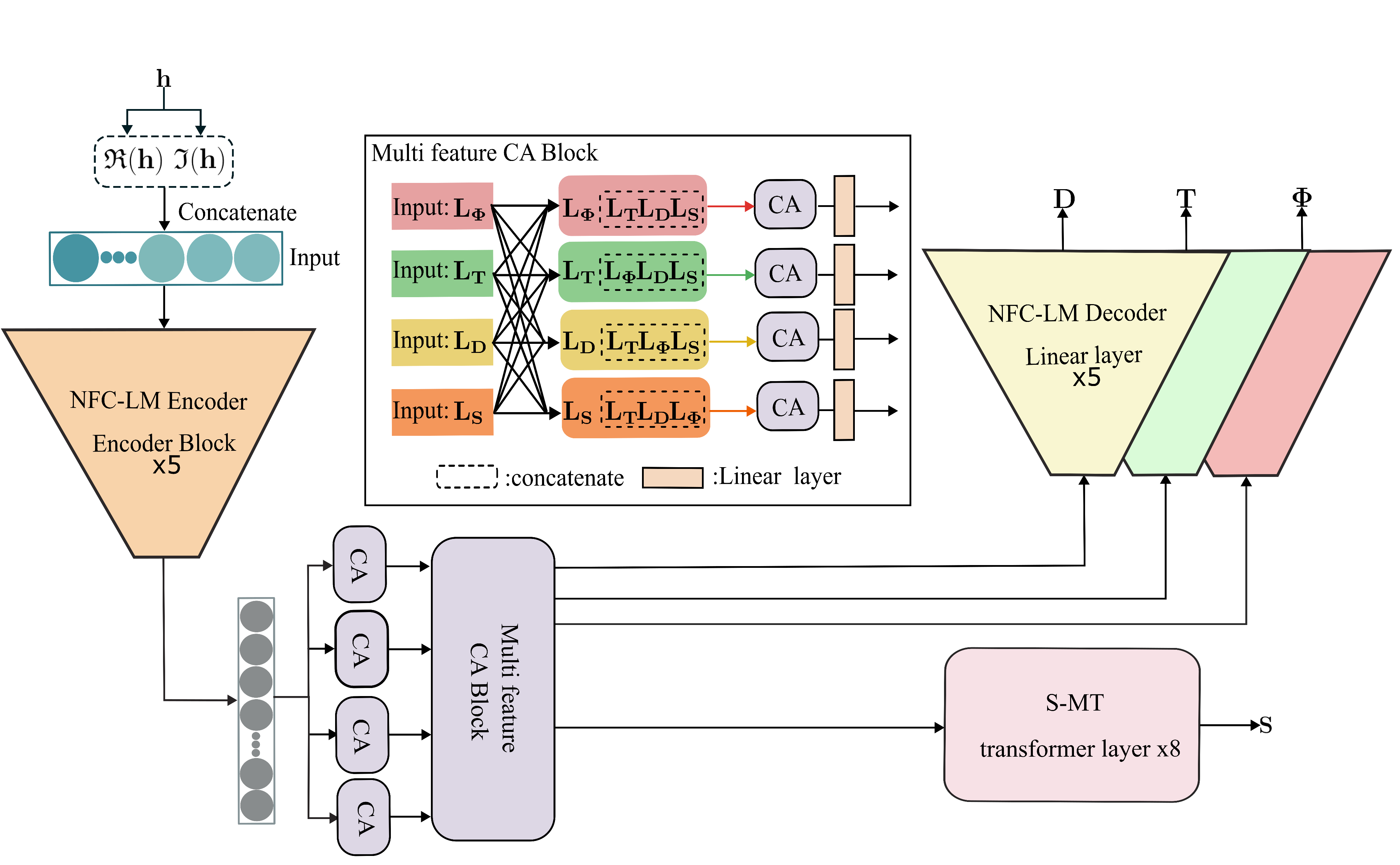}
\vspace{-0.4cm}
\caption{The network structure of the proposed adaptive TTD configuration beamforming method.}
\label{network_model}
\end{figure*}

\subsection{Network Architecture for Adaptive TTD Beamforming}

Our proposed network architecture for adaptive TTDs configuration in near-field beamforming includes the NFC-LM and the S-MT module. However, implementing a hard threshold-based strategy directly in the training of the entire network poses a challenge, particularly because the maximization operation \(\underset{\mathbf{P}}{\max} \sum_{i=1}^{n} \sum_{j=1}^{n} P_{ij} \cdot C_{ij}\) outlined in \textbf{Algorithm 1} can hinder the backpropagation process within the S-MT. This limitation makes it difficult to optimize the generated beamformers \(\mathbf{\Phi}\), \(\mathbf{T}\) and \(\mathbf{D}\) jointly with connection matrix \(\mathbf{S}\). To address this, we introduce a Multi-feature Channel Attention (MCA) block, denoted as \(\mathrm{MCA}(\cdot)\), to enhace the feature connections between the S-MT and NFC-LM modules. As showed in Fig. \ref{system_model}, each latent feature, characterized by beamformers and the switch network, is concatenated with others, serving as the \(\mathbf{Q}\).
% The whole network structure of our proposed adaptive TTD configuration for near-field beamforming consits of the NFC-LM and the S-MT module. However, it is challenging  to directly implement the hard threshold-based strategy in \(\underset{\mathbf{P}}{\max} \sum_{i=1}^{n} \sum_{j=1}^{n} P_{ij} \cdot C_{ij}\) in \textbf{Algorithm 1}
% the training of whole network, as \(\underset{\mathbf{P}}{\max} \sum_{i=1}^{n} \sum_{j=1}^{n} P_{ij} \cdot C_{ij}\) in \textbf{Algorithm 1} will stunt the backpropagation process of the NFC-LM. The generated beamformer \(\boldsymbol{\Phi}\), \(\boldsymbol{\mathcal{T}}\) and \(\boldsymbol{\mathcal{B}}\) will be hard for jointly optimizing with \(\mathbf{S}\).
% Therefore, we further introduce a multi feature CA block (\(\mathrm{MCA}(\cdot)\)) to enhance the feature connection between the S-MT and NFC-LM. As showed in Fig. \ref{system_model}, each feature stream, characterized by beamformers and the switch network, is interconnected with others, serving as the key value. 
After that, MCA block leverages cross-attention to facilitate the simultaneous optimization of switch selection and beamformer design. The \(\mathrm{MCA}(\cdot)\) processes inputs  \(\mathbf{L_{\Phi}}, \mathbf{L_{T}}, \mathbf{L_{D}} \) and \(\mathbf{L_{S}}\), producing the outputs as follows:
\vspace{-0.2cm}
\begin{subequations}
\begin{align}
\mathbf{Q} &= \mathrm{Linear}_{0,K+9}(\mathbf{X}')), \mathbf{K} = \mathrm{Linear}_{0,K+10} (\mathbf{X}), \\
\mathbf{V} &= \mathrm{Linear}_{0,K+11} (\mathbf{X}), \mathbf{X} = \mathrm{softmax} (\frac{\mathbf{Q}\mathbf{K}^{T}}{\sqrt{d_{k}}}) \cdot \mathbf{V},
\end{align}
\label{eq:QVVMCA}
\end{subequations} 
where \(\mathbf{X}' = \mathrm{con}(\mathbf{L_{\Phi}}, \mathbf{L_{T}}, \mathbf{L_{D}}, \mathbf{L_{S}})\) represents the concatenation of the inputs, and \(\mathbf{X} \in \{\mathbf{L_{\Phi}}, \mathbf{L_{T}}, \mathbf{L_{D}}, \mathbf{L_{S}}\}\) represents each input separately. Furthermore, considering the physical constraint that the TTD beamformers (\(\mathbf{T}\)) must be non-negative real numbers. 
Additionally, according to \cite{wang2023ttd}, the serial structure utilize increased time delays to improve beamforming performance. Thus, guaranteeing that each TTD provides a non-zero time delay can significantly enhance spectral efficiency. To maintain the backpropagation process while ensuring \(\mathbf{T}\) remains positive, we utilize the Softplus function, denoted as \(\mathrm{Softplus}(\cdot)\), rather than the ReLU function:
\vspace{-0.2cm}
\begin{equation}
\mathrm{Softplus}(x) = \ln(1 + e^x),
\label{eq:sft}
\vspace{-0.2cm}
\end{equation}
where \(x\) is the input to the \(\mathrm{Softplus}(\cdot)\). Different from the ReLU function, which outputs \(x\) for positive inputs and 0 otherwise, the Softplus function ensures the outputs are always positive and provides a differentiable gradient at every point, including \(x = 0\). Therefore, our proposed NFC-LM can adaptively optimze the time delay design of each TTD.

\subsection{Loss Function and Network Training Process}

We utilize the NFC-LM to design the \(\mathbf{\Phi}\), \(\mathbf{T}\) and \(\mathbf{D}\) through extracting the latent feature from the input channel response matrix \(\mathbf{H}\), and further utilize the S-MT to design the connection matrix \(\mathbf{S}\). While numerous DL-based hybrid beamforming methods transform the optimization problem into a convex problem and calculate the optimal result as training labels \cite{peken2020deep,ElbirDNN,elbir2019joint,vu2021machine}, this approach is impractical in our proposed physical structure. The proposed adaptive TTD configuration structure poses a significant challenge in providing optimal solutions for every possible connection scenarios. Therefore, we adopt an unsupervised, end-to-end training methodology for our hybrid beamforming network, eliminating the need for predetermined optimal solutions for  \({\mathbf{\Phi}}\), \({\mathbf{T}}\), \({\mathbf{D}}\) and \({\mathbf{S}}\).

The proposed loss function consists of two parts: the optimization objective and a set of regularization terms. The primary goal of the proposed adaptive TTD beamforming network is to maximize the spectral efficiency, as expressed in (\ref{eq:maxRm1}). To facilitate this, we first reshape the \({\mathbf{H}}\), \({\mathbf{\Phi}}\), \({\mathbf{T}}\), \({\mathbf{D}}\), and \({\mathbf{S}}\) into the same format \(\mathbb{R}^{K\times N_{R} \times M \times N}\). 
% Once reshaped, these can be expressed as \({\mathbf{H}}\),  \(\mathbf{\Phi}\), \(\mathbf{T}\), \(\mathbf{B}\), and \(\mathbf{\mathcal{S}}\), respectively. 
We then introduce an operation, \(\mathrm{H}_{+}\), which performs \(\odot\) across these matrices and sums the results over the \(N_{RF}\), \(M\), and \(N\) dimensions. The spectral efficiency optimization term for each \(k\) in \(K\) is expressed as
% The whole adaptive beamformer configuration network includes the wireless channel latent feature learning module, multi feature block and multi user transformer. To facilitate the training of whole network, the joint learning strategy is adopted. First, the wireless channel latent feature learning module is trained under fixed connection situation where \(C\) is a identity matrix. Then,  we train the whole network with multi feature block and multi user transformer for adaptive configuration by progressively interpolating the latent feature of different beamformers via the deep supervision of \(Loss_{Tmse}\), \(Loss_{Pt_{mse}}\) and \(Loss_{SE}\). The spectral efficiency loss \(Loss_{SE}\) is computed as
\vspace{-0.2cm}
\begin{equation}
\begin{split}
&\mathcal{L}_{\mathrm{Eff}} =  -\frac{1}{M+L_{\mathrm{CP}}} \times \\
&\log_2\left(1 + \frac{\left|\mathrm{H}_{+}(\mathbf{H}_{k},\mathbf{\Phi}_{k},  \mathbf{S}_{k} ,\mathbf{T}_{k},  \mathbf{D}_{k}) \right|^2}{\sum_{i=1, i \neq k}^{K} \left|\mathrm{H}_{+}(\mathbf{H}_{k}^H  ,\mathbf{\Phi}_{k} , \mathbf{S}_{k}, \mathbf{T}_{k} , \mathbf{D}_{k})\right|^2 + \sigma^2}\right).
\end{split}
\label{eq:leff}
\end{equation}
% We add three regularizers to the loss function to accelerate the training and ensure the rationality in real practice. We express the regularizers as follows:
To enhance the training process and ensure the practicality of our solutions, we introduce three regularization terms:
\begin{enumerate}
    \item \textbf{PS Modulus Constraint}: To maintain the constant modulus nature of the PSs, we employ the Mean Squared Error (MSE) on the magnitude of the predicted PS values \(\mathbf{\Phi}\) formulated as
    \vspace{-0.2cm}
    \begin{equation}
\mathcal{L}_{\mathrm{PS}} = \sum_{k=1}^{K} \sum_{N_{RF}=1}^{N_{R}} \sum_{m=1}^{M} \sum_{n=1}^{N} \left(|\mathbf{\Phi}_{kn_{r}mn}|^2 - 1 \right)^2.
        \label{eq:lps}
    \end{equation}
    \item \textbf{TTD Range Constraint}: To ensure that the predicted time delays are within the hardware's feasible range, we apply a conditional MSE that penalizes values outside this range, formulated as
    \vspace{-0.2cm}
\begin{equation}
\mathcal{L}_{\mathrm{TTD}} = \sum_{k=1}^{K} \sum_{N_{RF}=1}^{N_{R}} \sum_{m=1}^{M} \sum_{n=1}^{N} \psi(\mathbf{T}_{k,n_{r},m,n})
\vspace{-0.2cm}
\end{equation}
where
\begin{equation}
\vspace{-0.2cm}
\psi(x) = 
\begin{cases}
  (x - t_{\max})^2, & \text{if } x > t_{\max}, \\
  0, & \text{if } 0 < x < t_{\max}, \\
  x^2, & \text{if } x < 0.
\end{cases}
\end{equation}
    \item \textbf{Power Consumption Constraint:} To encourage energy-efficient beamforming and connection designs, we constrain the total power consumption to not exceed a predefined limit \(P_{t}\) expressed as
    \vspace{-0.2cm}
    \begin{equation}
        \mathcal{L}_{\mathrm{PC}} =  \left(\sum_{k=1}^{K}\sum_{N_{RF}=1}^{N_{R}} \sum_{m=1}^{M} \sum_{n=1}^{N}\left\| \mathbf{\Phi} \mathbf{S} \mathbf{T}  \mathbf{D}  \right\|_{F}^{2} - P_{t}\right)^2 .
        \label{eq:lpc}
        \vspace{-0.2cm}
    \end{equation}

\end{enumerate}
Combining these components, the total loss function for our adaptive beamformer training is formulated as
\vspace{-0.2cm}
\begin{equation}
\mathcal{L} = \mathcal{L}_{\mathrm{Eff}} + \mathcal{L}_{\mathrm{PS}} + \mathcal{L}_{\mathrm{TTD}} + \mathcal{L}_{\mathrm{PC}}.
\label{eq:ltotal}
\vspace{-0.2cm}
\end{equation}
This composite loss function is tailored to guide the network towards solutions that are not only optimal in terms of spectral efficiency but also adhere to practical constraints and hardware limitations.

.

\section{Numerical Results}
In this section, we conduct experiments to evaluate the effectiveness of our proposed deep unsupervised learning-based Adaptive TTD configuration for near-field beamforming.
% In this section, we conduct experiments 
% on the emulate ULA and UCA channel dataset under multi user scenario to evaluate our
% method and compare with other TTD configurations. 
\subsection{Simulation Setup}
\begin{table}[t!]
\centering
\caption{Simulation parameters.}
\vspace{-0.3cm}
\label{table:simulation_parameters}
\resizebox{0.45\textwidth}{!}{
\begin{tabular}{|l|c|}
\hline
\textbf{Parameter} & \textbf{Value} \\
\hline
Transmit power at the BS \( P_t \) & 20 dBm \\
\hline
Noise power density & \(-174\) dBm/Hz \\
\hline
Number of antennas at the BS \( N \) & 512 \\
\hline
System bandwidth \( B \) & 10 GHz \\
\hline
Central OFDM frequency \( f_c \) & 100 GHz \\
\hline
Number of OFDM subcarriers \( M \) & 10 \\
\hline
Length of OFDM cyclic prefix \( L_{CP} \) & 4 \\
\hline
Number of TTDs for each RF chain \( L \) & 32 \\
\hline
Maximum time delay of TTDs \( t_{max} \) & 80 ps \\
\hline
Number of channel paths \( L_k \) & 4 \\
\hline
Scattering loss \( \Lambda_\ell \) & \(-15\) dB \\
\hline
Radius of the UCA \(R\) & 0.768 m \\
\hline
Transmit and receive antenna gain \( G_t, G_r \) & 15 dB, 5 dB \\
\hline
\end{tabular}
}
\end{table}
To demonstrate the robustness of our proposed adaptive TTD configuration beamforming method across various antenna structures, we conducted experiments using two prevalent antenna configurations: the Uniform Linear Array (ULA) and the Uniform Circular Array (UCA).
% To demonstrate that our proposed method possesses better robustness to different antenna structures, we 
% select two common antenna structures Uniform Linear Array (ULA) and Uniform Circular Array (UCA) for the experiments.
For the ULA scenario, the scattering effects of the signal paths between the BS and users are numerically represented by \(L_{k}\). As expressed in (\ref{eq:hm,k}) and (\ref{brtheta}), the user positions are set by \(r_k\) and \(\theta_k\), representing the distance and direction relative to the center of the ULA at the BS, respectively. The position for the \(l\)-th scatter are denoted by \(\tilde{r}_{k,l}\) and \(\tilde{\theta}_{k,l}\). Here, \(C\) represents the speed of light, and \(\mathbf{r}(r,\theta) \in \mathbb{C}^{N \times 1} \) represents the distances wave travel from the BS antennas to the user. According to the near-field spherical wave model \cite{near-fieldYW}, the \(n\)-th component of \(\mathbf{r}(r,\theta)\), \([\mathbf{r}(r,\theta)]_{n}\) is defined as \(\sqrt{r^{2} + \delta^{2}_{n}d^{2}-2r\delta_{n}dcos\theta}\), with \(d = c/(2f_{c})\) indicating the antenna spacing and \(\delta_{n} \overset{\Delta}{=} n - 1 - \frac{N-1}{2} \). Similarly, for the UCA scenario, the positions of \(k\)-th user and \(l\)-th scatter are also represented by \( (r_k, \theta_k)\) and \( (\tilde{r}_{k,l}, \tilde{\theta}_{k,l}) \), respectively. Our model presumes the presence of user is confined to the two-dimensional plane coinciding with the UCA. Additionally, different from the ULA, the \(n\)-th component of \(\mathbf{r}(r,\theta)\) for UCA is formulated as \(\sqrt{r^{2} + R^{2} - 2rRcos(\theta -\psi_{n})}\), where \( R \) is is the UCA radius, and \( \psi_{n} = \frac{2\pi n}{N}\). To maintain consistent experimental conditions between the ULA and UCA, we calculate the radius such that \(2R = N \frac{C}{f_{c}}\).

For data generation, we assume users and scatters are randomly distributed in a semi-circular area spanning 0 to 180 degrees around the BS, with distances ranging from 5 to 15 meters. User positions are precisely sampled at 0.1-meter intervals and 0.5-degree angles within this domain. Following the (\ref{eq:hm,k}), we generated 10,000 datasets for both ULA and UCA channels, with each user associated with four scatterers.
% Consequently, we apply an identical sampling method to create Non-Line-of-Sight (NLOS) channel datasets concurrently. 
Consequently, every channel dataset is a blend of LOS and NLOS channel information. The datasets were divided into training (60\%), testing (20\%), and validation (20\%) sets.
% The number of RF chains is set to \(N_{RF} = \) 4 for multi user systems.

In the millimeter-wave and terahertz bands, path loss is mainly affected by propagation loss, but absorption loss can also have an effect. The path loss for a given frequency \(f\) and propagation distance \(r\) can be expressed as 
\vspace{-0.2cm}
\begin{equation}
    \eta_{pathloss}(f,r) = \left(\frac{4\pi fr}{c}\right)^{2}e^{k_{abs}(f)r},
    \vspace{-0.2cm}
\end{equation}
where \(k_{abs}\) denotes the frequency-specific medium absorption coefficient, obtainable from the HITRAN database \cite{gordon2017hitran2016}. The LOS channel gain is formulated as \( \left\| \beta_{m,k} \right\|^{2} = \eta^{-1}_{pathloss}(f_{m},r_{k})G_{r}G_{t}\), with \(G_{r}\) and \(G_{t}\) receiver and transmitter antenna gains, respectively. For the NLoS component, incorporating scattering loss is crucial. The gain from the \(l\)-th NLoS component is expressed as \( \left\| \beta_{m,k,l} \right\|^{2} = \Lambda _{l}\eta^{-1}_{pathloss}(f_{m},r_{k})G_{r}G_{t}\), where \(\Lambda _{l}\) denotes the scattering loss and \(r_{k,l}\) denotes the propagation distance to user \(k\) through the \(l\)-th scatterer. Unless specified otherwise, the simulation  follows parameters listed in Table I.

Our method was implemented in Pytorch and executed on an NVIDIA GeForce A40 GPU. During the training phase, we utilized a batch size of 2 and employed the Adam optimizer across 1000 epochs. 
% The learning rate was initially set to \(1\times10^{-5}\) and decaying each 10 epochs with factor 0.5. To avoid overfitting, we adopted an early stopping mechanism when the validation loss is not reduced more than 50 epochs. 
% Training the whole network takes about 30 minutes.

For comparison, we consider the following three benchmark scenarios:
\begin{itemize}
    \item \textbf{Optimal Full-Digital Beamforming (BF)} :In this scenario, each antenna is individually connected to a dedicated Radio Frequency (RF) chain, enabling the creation of a comprehensive baseband digital beamformer for every subcarrier. This setup is designed to establish a theoretical upper bound for performance metrics.
    \item \textbf{Optimal Time Delay Beamforming (TTD-BF)} :This benchmark presumes the availability of TTDs with an infinite range, i.e.,\(t_{max} = +\infty\), thereby facilitating the highest possible performance for the TTD-BF architecture.
    \item \textbf{Conventional Beamforming (CB)}: This benchmark corresponds to a Phase Shifter (PS)-only hybrid BF architecture. It is constrained to frequency-independent analog beamforming capabilities.
\end{itemize}
\begin{figure}[t!]
\centering
\includegraphics[width=0.48\textwidth]{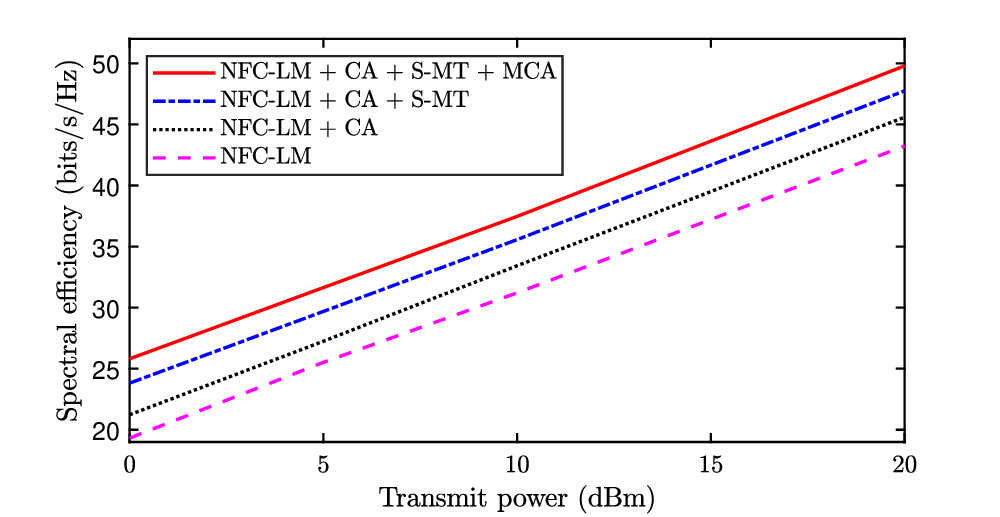}
  \vspace{-0.4cm}
 \caption{Spectral efficiency results of different network structures based on the NFC-LM across varying levels of transmit power.}
 \label{img:ab1}
\end{figure}
\begin{table}[t!]
\centering
% \captionsetup{justification=centering, labelsep=newline}
\vspace{-0.3cm}
\caption{Ablation Study on the Number of Transformer Layers (\(I\)) and Embedding Feature Dimensionality}
\vspace{-0.3cm}
\label{table:ablation_study}
\resizebox{\columnwidth}{!}{%
\begin{tabular}{cc|ccccc}
\hline
Layer & Feature & \multicolumn{5}{c}{Spectral Efficiency (bit/s/Hz)} \\
\cline{3-7}
 \# (\(I\)) & Dim. & 0 dBm & 5 dBm & 10 dBm & 15 dBm & 20 dBm \\
\hline
1 & 512 & 22.47 & 29.26 & 33.89 & 40.27 & 46.83 \\
8 & 512 & 24.31 & 30.31 & 36.00 & 42.09 & 48.25 \\
8 & 512 & \textbf{25.81} & \textbf{31.64} & \textbf{37.47} & \textbf{43.63} & \textbf{49.79} \\
4 & 256 & 23.81 & 29.69 & 35.57 & 41.76 & 47.79 \\
4 & 768 & 25.28 & 31.29 & 36.84 & 43.12 & 49.19 \\
\hline
\end{tabular}%
}
\end{table}

\subsection{Ablation Studies}
\begin{figure}[t!]
    \centering
    \begin{minipage}[b]{0.48\textwidth}
        \includegraphics[width=\textwidth]{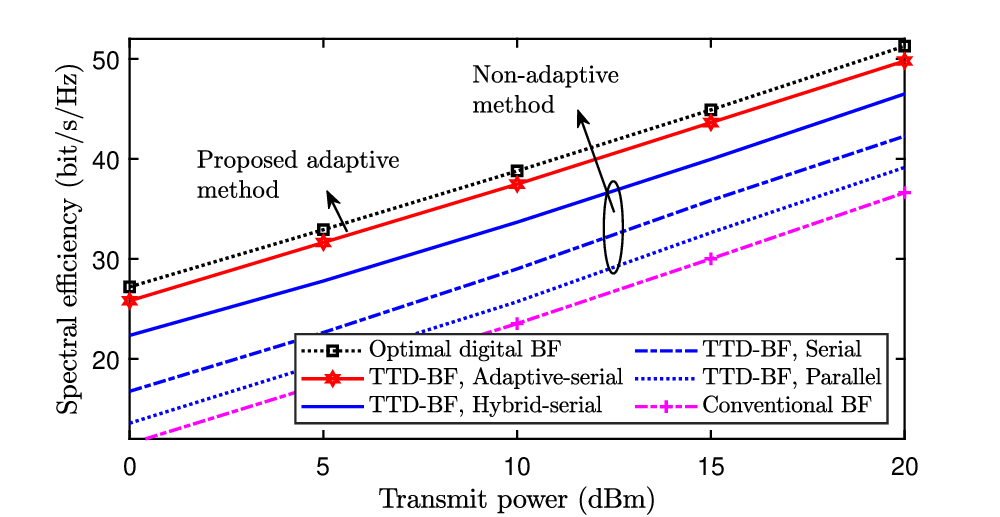}
        \vspace{-0.9cm}
        \caption{Average spectral efficiency versus the maximum transmit
power in the multi-user ULA system.}
        \label{img:ULARTP}
    \end{minipage}
    \hfill
    \begin{minipage}[t!]{0.48\textwidth}
        \includegraphics[width=\textwidth]{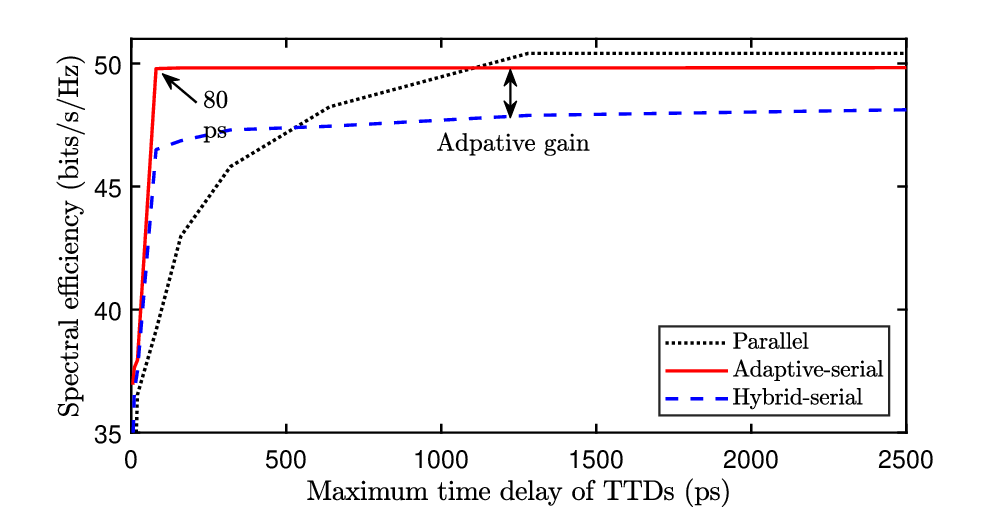}
        \vspace{-0.9cm}
        \caption{Average spectral efficiency versus the maximum time delay \(t_{max}\) of TTDs in the multi-user ULA system.}
        \label{img:ULARTTD}
    \end{minipage}
\end{figure}

% \begin{figure} [t!]
% \centering
% \includegraphics[width=0.48\textwidth]{TTD/MULTI_R_values_plot.eps}
% \vspace{-0.9cm}
%  \caption{ Average spectral efficiency versus the maximum transmit
% power in the multi-user ULA system.
%   }
%  \label{img:ULARTP}
% \end{figure}
% \begin{figure} [t!]
% \vspace{-0.6cm}
% \centering
% \includegraphics[width=0.48\textwidth]{TTD/TTD_ULA.eps}
% \vspace{-0.9cm}
%  \caption{Average spectral efficiency versus the maximum time delay \(t_{max}\) of TTDs in the multi-user ULA system.
%   }
%  \label{img:ULARTTD}
% \end{figure}
In this subsection, we conduct a series of ablation studies to evaluate the effect of critical components and hyperparameter settings of our proposed adaptive TTD configuration beamforming method on spectral efficiency, particularly for short-range TTDs. These essential components include the near-field channel learning module (NFC-LM), the cross attention (CA) module, the switch multi-user transformer (S-MT), and the multi-feature cross attention (MCA) module. The generated ULA dataset is used in the ablation studies.
We treat the NFC-LM as our baseline model which consists of multiple linear decoders and a convolutional encoder but excludes the CA, S-MT, and MCA modules. Moreover, to demonstrate the effectiveness of our proposed S-MT, the NFC-LM employs an additional linear decoder in place of the S-MT which is identical to those decoding the variables \(\mathbf{\Phi}\), \(\mathbf{T}\) and \(\mathbf{D}\). Despite these modifications, we continue to apply the Hungarian algorithm to map the predicted connection matrix \(\mathbf{S}'\) to a binary matrix \(\mathbf{S}\) . The comparisons of spectral efficiency results are shown in Fig. \ref{img:ab1} and Table II.

% Moreover, for comparison, our baseline method is executed using multiple linear decoders and one convolutional channel feature encoder but without the CA module, S-MT module and the MCA module. Furthermore, in order to demonstrate the efficiency of our proposed S-MT, in our baseline method, we replace the S-MT with a linear decoder. This decoder is same as the one utilized for decoding other variables \(\mathbf{\Phi}\), \(\mathbf{T}\) and \(\mathbf{B}\). And we still utilize Hungarian algorithm to project the output connection matrix \(\mathbf{\mathcal{S}}\) to a binary connection matrix. The comparisons of spectral efficiency results are shown in Fig. \ref{img:ab1} and Table II. 

As shown in Fig. \ref{img:ab1}, we gradually incorporate the CA, S-MT, and MCA modules into the NFC-LM across various transmit power settings, denoted as “NFC-LM”, “NFC-LM + CA”, “NFC-LM + CA + S-MT”, and “NFC-LM + CA + S-MT + MCA”, with the last configuration representing our proposed method.
% conduct ablation studies of our method by gradually adding the CA, S-MT, and MCA to the baseline network on different Transmit powers which are denoted as: 'baseline', 'baseline + CA', 'baseline + CA + S-MT', and 'baseline + CA + S-MT + MCA', with the final configuration representing our fully enhanced method. 
The results clearly indicate a progressive improvement in spectral efficiency with the sequential integration of CA, S-MT and MCA. The NFC-LM, operating with short-range TTDs, shows the lower bound of our proposed latent feature learning based adaptive TTD beamforming method. Additionally, our fully enhanced method, 'NFC-LM + CA + S-MT + MCA',  achieves an average improvement of approximately \( 6.5 \) bit/s/Hz in the results over the NFC-LM across varying levels of transmit power. The CA module, designed to establish the correlation between the near-field channel latent feature \(\mathbf{H}\) and the latent features of multiple beamformers \(\mathbf{\Phi}\), \(\mathbf{T}\) and \(\mathbf{D}\), augments spectral efficiency by more than 1 bit/s/Hz on average compared to the NFC-LM. 
% Additionally, it facilitates the correlation between \(\mathbf{H}\) and the connection matrix \(\mathbf{\mathcal{S}}\), respectively. The results show that the spectral efficiency of CA increases more than \( 1 \) in average than the baseline network. 
Moreover, the S-MT is capable of optimizing the adaptive connection of the TTD through multi-user attention (MSA) and positional coding (PC), effectively enhancing the performance of the network. In addition, the MCA module further enhances the joint optimization of \(\mathbf{\Phi}\), \(\mathbf{T}\), \(\mathbf{D}\) and \(\mathbf{S}\), leading to an average spectral efficiency increase of 2 bit/s/Hz.
% \begin{figure} [t!]
% \centering
% \includegraphics[width=0.48\textwidth]{TTD/MULTI_R_values_plot_UCA.eps}
%  \vspace{-0.9cm}
%  \caption{Average spectral efficiency versus the maximum transmit
% power in the multi-user UCA system.
%   }
%  \label{img:UCARTP}
% \end{figure}
% \begin{figure} [t!]
% \vspace{-0.5cm}
% \centering
% \includegraphics[width=0.48\textwidth]{TTD/TTD_UCA.eps}
% \vspace{-0.9cm}
%  \caption{Average spectral efficiency versus the maximum time delay \(t_{max}\) of TTDs in the multi-user UCA system.
%   }
%  \label{img:UCARTTD}
% \end{figure}
\begin{figure}[t!]
    \centering
    \begin{minipage}[b]{0.48\textwidth}    \includegraphics[width=\textwidth]{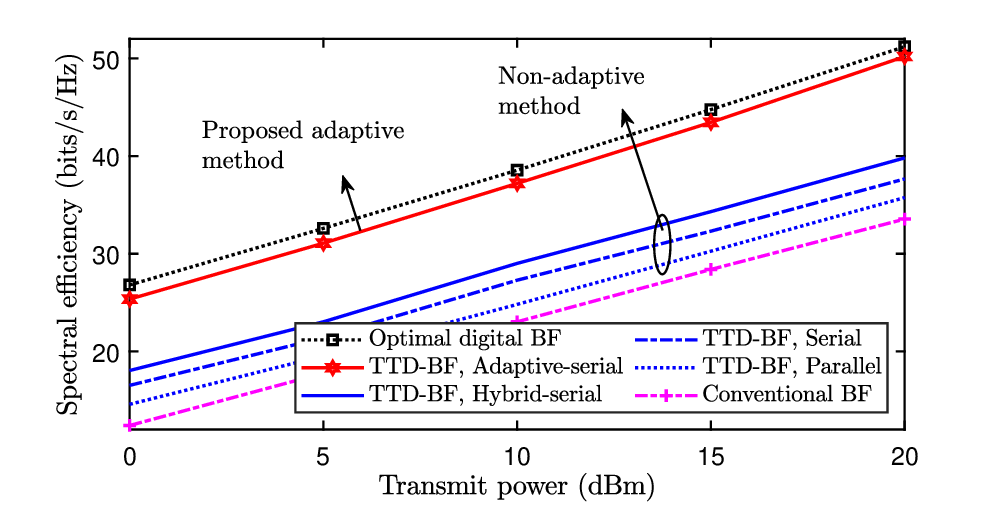}
    \vspace{-0.9cm}
        \caption{Average spectral efficiency versus the maximum transmit
        power in the multi-user UCA system.}
        \label{img:UCARTP}
    \end{minipage}
    \hfill
    \begin{minipage}[t!]{0.48\textwidth}
    \includegraphics[width=\textwidth]{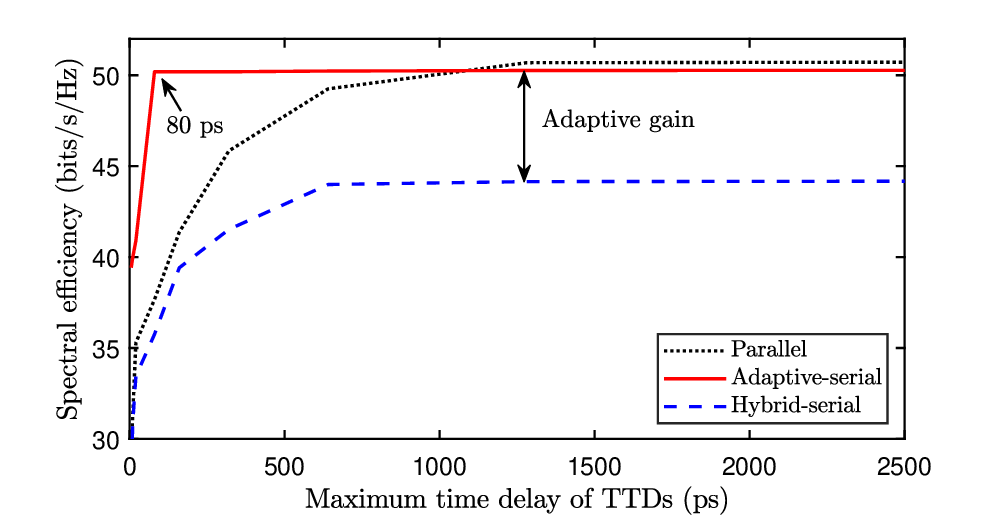}
    \vspace{-0.9cm}
        \caption{Average spectral efficiency versus the maximum time delay \(t_{max}\) of TTDs in the multi-user UCA system.}
        \label{img:UCARTTD}
    \end{minipage}
\end{figure}

Furthermore, we conduct ablation experiments to evaluate the impact of hyperparameter configurations of S-MT on spectral efficiency. We specifically examine the symmetrical structure of the transformer by modifying the number of layers (\(I\)) at {1, 4, 8}, and by modifying the feature dimensionality in the linear projection layer of the Feedforward Network (FFN) to {256, 512, 768}. The comparative analysis, presented in Table I, clearly shows the spectral efficiency at different transmit power levels. We can find that the optimal performance is obtained with a network configuration of \(I = 8\) layers and a feature dimensionality of 512, yielding spectral efficiency of 28.82, 34.6, 37.97, 42.06 and 47.95 bit/s/Hz at transmit powers of 0, 5, 10, 15, and 20 dBm, respectively.

% \begin{table}[t!]
% \centering
% % \captionsetup{justification=centering, labelsep=newline}
% \caption{Ablation Study on the Number of Transformer Layers (\(I\)) and Embedding Feature Dimensionality}
% \label{table:ablation_study}
% \resizebox{\columnwidth}{!}{%
% \begin{tabular}{cc|ccccc}
% \hline
% Layer & Feature & \multicolumn{5}{c}{Spectral Efficiency (bit/s/Hz)} \\
% \cline{3-7}
%  \# (\(I\)) & Dim. & 0 dBm & 5 dBm & 10 dBm & 15 dBm & 20 dBm \\
% \hline
% 1 & 512 & 22.47 & 29.26 & 33.89 & 40.27 & 46.83 \\
% 8 & 512 & 24.31 & 30.31 & 36.00 & 42.09 & 48.25 \\
% 8 & 512 & \textbf{25.81} & \textbf{31.64} & \textbf{37.47} & \textbf{43.63} & \textbf{49.79} \\
% 4 & 256 & 23.81 & 29.69 & 35.57 & 41.76 & 47.79 \\
% 4 & 768 & 25.28 & 31.29 & 36.84 & 43.12 & 49.19 \\
% \hline
% \end{tabular}%
% }
% \end{table}

\subsection{Spectral Efficiency versus Transmit Power}

In our analysis of multi-user scenarios, we explore the spectral efficiency of ULA and UCA systems under conditions of maximum transmit power and diverse TTD configurations. The evaluations, as presented in Fig. \ref{img:ULARTP} and Fig. \ref{img:UCARTP} are based on scenarios where the distance between users and the BS varies randomly from 5 to 15 meters. Our experimental setup involves \(K\) = 4 users, \(N_{RF}\) = 4 RF chains, and a TTD contraint of \(t_{max}\) = 80 ps. Here, “Hybrid” implies that the TTDs for all RF chains are cascaded in serial configuration and hybrid connected with the PSs \cite{ZhaoTTD}. And the "parallel" means that each TTD is connected to an individual antenna. As showed in Fig. \ref{img:ULARTP}, under the ULA scenario, the CB configuration achieves comparable results to the parallel TTD setup within the given time delay constraints. It is notable that serial configurations surpass their parallel counterparts in performance, attributed to their superior capability in providing more extensive time delay with short-range TTDs. Furthermore, the hybrid configuration, which optimizes the effective coverage area, demonstrates a significant improvement in performance. Therefore, when employing an adaptive TTD configuration which optimizes the connection between TTDs and PSs, our proposed method approaches the theoretical optimum. In addition, thanks to the adaptive TTD configuration, our proposed method can have less performance degradation under different antenna structures. As showed in Fig. \ref{img:UCARTP}, our proposed method still achieves comparable results to the theoretical optimum, however, hybrid configuration experience severe performance degradation. Changing the antenna structure disrupts the guaranteed monotonic correspondence between time delays and user positions \cite{wang2023ttd}. Thus, we can find that the parallel configuration outperforms the serial configuration. The results suggest that our proposed adaptive TTD configurations can substantially enhance spectral efficiency in both multi-user ULA and UCA systems.

% \begin{figure} [t!]
% \centering
% \includegraphics[width=0.48\textwidth]{TTD/MULTI_R_values_plot_UCA.eps}
%  \vspace{-0.7cm}
%  \caption{Average spectral efficiency versus the maximum transmit
% power in the multi-user UCA system.
%   }
%  \label{img:UCARTP}
% \end{figure}

% \begin{figure} [t!]
% \centering
% \includegraphics[width=0.48\textwidth]{TTD/CDF_ULA.eps}
% \vspace{-0.5cm}
%  \caption{CDF versus average spectral efficiency for multi-user ULA system.
%   }
%  \label{img:ULACDF}
% \end{figure}
\begin{figure} [t!]
\centering
\includegraphics[width=0.48\textwidth]{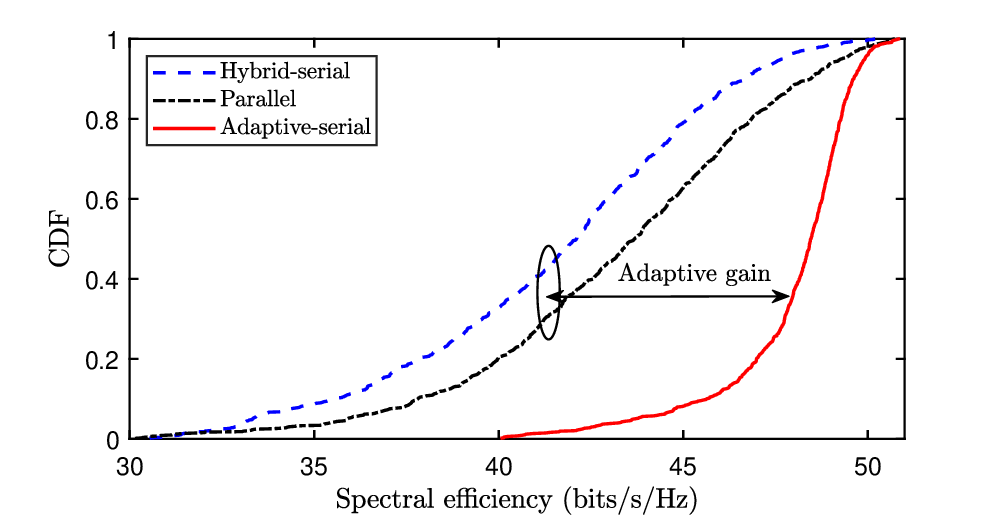}
\vspace{-0.4cm}
 \caption{CDF versus average spectral efficiency for multi-user ULA system.
  }
 \label{img:ULACDF}
\end{figure}
\begin{figure} [t!]
\vspace{-0.4cm}
\centering
\includegraphics[width=0.48\textwidth]{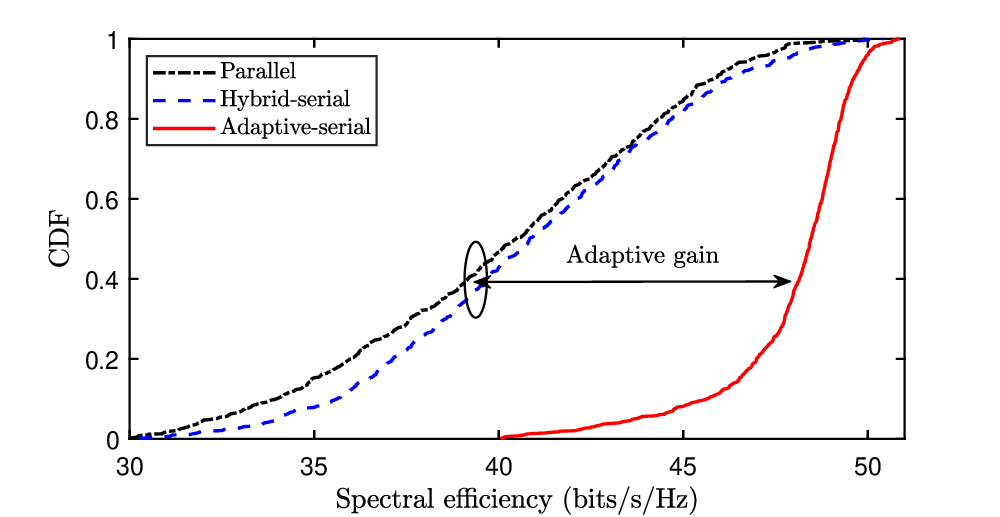}
\vspace{-0.4cm}
 \caption{CDF versus average spectral efficiency for multi-user UCA system.}
 \label{img:UCACDF}
\end{figure}
\subsection{Spectral Efficiency versus Maximum Time Delay}

% \begin{figure} [t!]
% \centering
% \includegraphics[width=0.48\textwidth]{TTD/TTD_ULA.eps}
% \vspace{-0.7cm}
%  \caption{Average spectral efficiency versus the maximum time delay \(t_{max}\) of TTDs in the multi-user ULA system.
%   }
%  \label{img:ULARTTD}
% \end{figure}

% \begin{figure} [t!]
% \centering
% \includegraphics[width=0.48\textwidth]{TTD/TTD_UCA.eps}
% \vspace{-0.73cm}
%  \caption{Average spectral efficiency versus the maximum time delay \(t_{max}\) of TTDs in the multi-user UCA system.
%   }
%  \label{img:UCARTTD}
% \end{figure}

To further illustrate the effectiveness of our proposed adaptive TTD configurations, we examine the relationship between spectral efficiency and infinite maximum time delay \(t_{max}\) fot TTDs, as illustrated in Fig. \ref{img:ULARTTD} and Fig. \ref{img:UCARTTD}. Particularly for maximum time delays below 80 ps, the adaptive configuration shows superior performance compared to the other configurations. This indicates that the adaptive approach effectively leverages dynamic switching to mitigate time delay discrepancies among users, thereby sustaining higher spectral efficiency even with limited time delay capacity. This observation demonstrates the robustness of our proposed adaptive TTD configuration in environments constrained by finite and narrower TTD ranges. As \(t_{max}\) increases, all configurations show an clear increase in spectral efficiency, but they exhibit different behaviors. In the ULA system, as shown in Fig. \ref{img:ULARTTD}, the parallel configuration necessitates a \(t_{max}\) of at least 500 ps to match the performance of the serial configuration. However, as shown in Fig. \ref{img:UCARTTD}, within the UCA system, parallel configuration obviously outperforms the serial configuration in pectral efficiency. For the UCA system, achieving comparable performance with the ULA system through a serial configuration demands a significantly high time delay, surpassing 500 ps. Moreover, we can find that our proposed adaptive configuration achieves comparable results to the parallel configuration when considering an infinite time delay. However, as the maximum time delay \(t_{max}\) increases, the parallel configuration begins to show increasing benefits. A primary advantage of the parallel configuration  is its direct compensation for each antenna subarray by its respective TTD, simplifying the computational demands. Therefore, for applications accommodating large time delays while aiming to minimize hardware complexity, the parallel configuration may offer enhanced performance and increased adaptability.

% \begin{figure} [t!]
% \centering
% \includegraphics[width=0.48\textwidth]{TTD/TTD_ULA.eps}
% \vspace{-0.7cm}
%  \caption{Average spectral efficiency versus the maximum time delay \(t_{max}\) of TTDs in the multi-user ULA system.
%   }
%  \label{img:ULARTTD}
% \end{figure}

% \begin{figure} [t!]
% \centering
% \includegraphics[width=0.48\textwidth]{TTD/TTD_UCA.eps}
% \vspace{-0.7cm}
%  \caption{Average spectral efficiency versus the maximum time delay \(t_{max}\) of TTDs in the multi-user UCA system.
%   }
%  \label{img:UCARTTD}
% \end{figure}

% \begin{figure} [t!]
% \centering
% \includegraphics[width=0.48\textwidth]{TTD/CDF_ULA.eps}
% \vspace{-0.5cm}
%  \caption{CDF versus average spectral efficiency for multi-user ULA system.
%   }
%  \label{img:ULACDF}
% \end{figure}

% \begin{figure} [t!]
% \centering
% \includegraphics[width=0.48\textwidth]{TTD/CDF_UCA.eps}
% \vspace{-0.9cm}
%  \caption{CDF versus average spectral efficiency for multi-user UCA system.}
%  \label{img:UCACDF}
% \end{figure}
% \subsection{Cumulative Distribution of Spectral Efficiency with Different TTD Configurations}
\subsection{Cumulative Distribution of Spectral Efficiency}
To demonstrate the robustness of our proposed adaptive TTD configuration beamforming in near-field region, we present Figures \ref{img:ULACDF} and \ref{img:UCACDF}. For these analyses, we fix power consumption at 20 dB and the time delay at 80 ps. We position the users uniformly at a distance of 10 meters from the BS, with their angular distribution ranging from 0 to 180 degrees. 
% Moreover, we present Fig. \ref{img:ULACDF} and Fig. \ref{img:UCACDF} to illustrate the robustness of our proposed TTD adaptive configuration hybrid beamforming method. We first fix the power consumption and the time delay as 20db and 80ps, respectively. And we set the user in the same distance as 10m away from the BS and randomly distributed between 0 to 180 degrees. 
As showed in Fig. \ref{img:ULACDF}, we can find that the serial and parallel configuration have similar distribution and serial has higher spectral efficiency. This adavantage stems from the serial configuration can provide cumulative time delay compensation, which helps alleviate the spatial-wideband effect with short-range TTDs. Furthermore, the spectral efficiency of our proposed adaptive TTD configuration mainly distributed between 40 and 56, indicating a performance enhancement over the serial configuration. Similarly, under the UCA system, as shown in Figure \ref{img:UCACDF}, our proposed adaptive configuration surpasses both the parallel and serial configurations, with spectral efficiency values distributed between 40 and 53. This is beacuse the adaptive configuration are more robust and can alleviate the performance decrease under different antenna systems. Additionally, it is observed that, under the UCA system, the parallel configuration achieves performance comparable to the serial configuration.

\section{Conclusion}
In this article, we introduce an adaptive TTD configuration for short-range TTDs. Compared to other existing TTD configurations, our proposed method can effectively combat the spatial-wideband effect for arbitrary user locations and array shape by dynamically selecting the connections between TTDs and PSs. We proposed a novel end-to-end deep neural network which consists of near-field channel learning (NFC-LM) module and switch multi-user transformer (S-MT) module. With encoding-decoding structure, NFC-LM explores the latent feature of near-field channel response and utilzes cross attention (CA) module to construct the relations among channel response, analog beamformers, digital baemformer and switch network. Moreover, the S-MT guides the connection between TTDs and antenna subarrays through multi-user attention (MSA) and positional coding (PC). Furthermore, Multi feature Cross Attention module (MCA) is proposed to promote the joint optimization of beamformer design and switch network configuration. Simulation results demonstrates that our proposed adaptive TTD configuration outperforms other TTD configuration hybrid beamforming methods across various antenna structures.

\bibliographystyle{IEEEtran}
\bibliography{mybib}

\end{document}